\documentclass[longauth]{aa}

\usepackage{longtable,lscape}
\usepackage{txfonts}
\usepackage{graphicx}
\usepackage{url,hyperref}
\usepackage{dcolumn} 
\usepackage{natbib}
\usepackage{ulem}
\usepackage{color}
\usepackage[flushleft]{threeparttable}
\usepackage[T1]{fontenc}

\usepackage{comment}
\bibpunct{(}{)}{;}{a}{}{,} 

\newcommand{\HMS}[3]{$#1\mbox{$^{\mathrm h}$}\,#2\mbox{$^{\mathrm m}$}\,#3\mbox{$^{\mathrm s}$}$}
\newcommand{\DMS}[3]{$#1\mbox{$^\circ$}\,#2\arcmin\,#3\arcsec$}

\let\linenumbers\relax
\let\nolinenumbers\relax

\begin{document}


\title{The H.E.S.S. extragalactic sky survey with the first decade of observations}
\titlerunning{HEGS : The H.E.S.S. extragalactic survey}

\author{H.E.S.S. Collaboration
\and F.~Aharonian \inst{\ref{DIAS},\ref{MPIK},\ref{Yerevan}}
\and F.~Ait~Benkhali \inst{\ref{LSW}}
\and J.~Aschersleben \inst{\ref{Groningen}}
\and H.~Ashkar \inst{\ref{LLR}}
\and M.~Backes \inst{\ref{UNAM},\ref{NWU}}
\and V.~Barbosa~Martins \inst{\ref{DESY}}
\and R.~Batzofin \inst{\ref{UP}}
\and Y.~Becherini \inst{\ref{APC},\ref{Linnaeus}}
\and D.~Berge \inst{\ref{DESY},\ref{HUB}}
\and K.~Bernl\"ohr \inst{\ref{MPIK}}
\and M.~B\"ottcher \inst{\ref{NWU}}
\and C.~Boisson \inst{\ref{LUTH}}
\and J.~Bolmont \inst{\ref{LPNHE}}
\and M.~de~Bony~de~Lavergne \inst{\ref{CEA}}
\and J.~Borowska \inst{\ref{HUB}}
\and M.~Bouyahiaoui \inst{\ref{MPIK}}
\and F.~Bradascio \inst{\ref{CEA}}
\and R.~Brose \inst{\ref{DIAS}}
\and A.~Brown \inst{\ref{Oxford}}
\and F.~Brun\protect\footnotemark[1] \inst{\ref{CEA}}
\and B.~Bruno \inst{\ref{ECAP}}
\and T.~Bulik \inst{\ref{UWarsaw}}
\and C.~Burger-Scheidlin \inst{\ref{DIAS}}
\and T.~Bylund \inst{\ref{CEA}}
\and S.~Casanova \inst{\ref{IFJPAN}}
\and J.~Celic \inst{\ref{ECAP}}
\and M.~Cerruti \inst{\ref{APC}}
\and T.~Chand \inst{\ref{NWU}}
\and S.~Chandra \inst{\ref{NWU}}
\and A.~Chen \inst{\ref{Wits}}
\and J.~Chibueze \inst{\ref{NWU}}
\and O.~Chibueze \inst{\ref{NWU}}
\and G.~Cotter \inst{\ref{Oxford}}
\and P.~Cristofari \inst{\ref{LUTH}}
\and J.~Damascene~Mbarubucyeye \inst{\ref{DESY}}
\and I.D.~Davids \inst{\ref{UNAM}}
\and J.~Devin \inst{\ref{LUPM}}
\and J.~Djuvsland \inst{\ref{MPIK}}
\and A.~Dmytriiev \inst{\ref{NWU}}
\and K.~Egberts \inst{\ref{UP}}
\and S.~Einecke \inst{\ref{Adelaide}}
\and S.~Fegan \inst{\ref{LLR}}
\and G.~Fontaine \inst{\ref{LLR}}
\and S.~Funk \inst{\ref{ECAP}}
\and S.~Gabici \inst{\ref{APC}}
\and J.F.~Glicenstein \inst{\ref{CEA}}
\and J.~Glombitza \inst{\ref{ECAP}}
\and P.~Goswami \inst{\ref{APC}}
\and G.~Grolleron \inst{\ref{LPNHE}}
\and L.~Haerer \inst{\ref{MPIK}}
\and B.~He\ss \inst{\ref{IAAT}}
\and J.A.~Hinton \inst{\ref{MPIK}}
\and W.~Hofmann \inst{\ref{MPIK}}
\and T.~L.~Holch \inst{\ref{DESY}}
\and M.~Holler \inst{\ref{Innsbruck}}
\and D.~Horns \inst{\ref{UHH}}
\and Zhiqiu~Huang \inst{\ref{MPIK}}
\and M.~Jamrozy \inst{\ref{UJK}}
\and F.~Jankowsky \inst{\ref{LSW}}
\and A.~Jardin-Blicq \inst{\ref{CENBG}}
\and E.~Kasai \inst{\ref{UNAM}}
\and K.~Katarzy{\'n}ski \inst{\ref{NCUT}}
\and R.~Khatoon \inst{\ref{NWU}}
\and B.~Kh\'elifi \inst{\ref{APC}}
\and Nu.~Komin \inst{\ref{Wits}}
\and K.~Kosack \inst{\ref{CEA}}
\and D.~Kostunin \inst{\ref{DESY}}
\and A.~Kundu \inst{\ref{NWU}}
\and R.G.~Lang \inst{\ref{ECAP}}
\and S.~Le~Stum \inst{\ref{CPPM}}
\and A.~Lemi\`ere \inst{\ref{APC}}
\and M.~Lemoine-Goumard \inst{\ref{CENBG}}
\and J.-P.~Lenain \inst{\ref{LPNHE}}
\and F.~Leuschner \inst{\ref{IAAT}}
\and A.~Luashvili \inst{\ref{LUTH}}
\and J.~Mackey \inst{\ref{DIAS}}
\and D.~Malyshev \inst{\ref{IAAT}}
\and V.~Marandon \inst{\ref{CEA}}
\and G.~Mart\'i-Devesa \inst{\ref{Innsbruck}}
\and R.~Marx \inst{\ref{LSW}}
\and A.~Mehta \inst{\ref{DESY}}
\and A.~Mitchell \inst{\ref{ECAP}}
\and R.~Moderski \inst{\ref{NCAC}}
\and L.~Mohrmann \inst{\ref{MPIK}}
\and A.~Montanari \inst{\ref{LSW}}
\and M.~de~Naurois \inst{\ref{LLR}}
\and J.~Niemiec \inst{\ref{IFJPAN}}
\and P.~O'Brien \inst{\ref{Leicester}}
\and L.~Olivera-Nieto \inst{\ref{MPIK}}
\and E.~de~Ona~Wilhelmi \inst{\ref{DESY}}
\and M.~Ostrowski \inst{\ref{UJK}}
\and S.~Panny \inst{\ref{Innsbruck}}
\and M.~Panter \inst{\ref{MPIK}}
\and U.~Pensec \inst{\ref{LPNHE}}
\and G.~P\"uhlhofer \inst{\ref{IAAT}}
\and M.~Punch \inst{\ref{APC}}
\and A.~Quirrenbach \inst{\ref{LSW}}
\and S.~Ravikularaman \inst{\ref{APC},\ref{MPIK}}
\and M.~Regeard \inst{\ref{APC}}
\and A.~Reimer \inst{\ref{Innsbruck}}
\and O.~Reimer \inst{\ref{Innsbruck}}
\and I.~Reis \inst{\ref{CEA}}
\and H.~Ren \inst{\ref{MPIK}}
\and B.~Reville \inst{\ref{MPIK}}
\and F.~Rieger \inst{\ref{MPIK}}
\and G.~Rowell \inst{\ref{Adelaide}}
\and B.~Rudak \inst{\ref{NCAC}}
\and E.~Ruiz-Velasco \inst{\ref{LAPP}}
\and V.~Sahakian \inst{\ref{Yerevan2}}
\and H.~Salzmann \inst{\ref{IAAT}}
\and D.A.~Sanchez\protect\footnotemark[1] \inst{\ref{LAPP}}
\and A.~Santangelo \inst{\ref{IAAT}}
\and M.~Sasaki \inst{\ref{ECAP}}
\and J.~Sch\"afer \inst{\ref{ECAP}}
\and F.~Sch\"ussler \inst{\ref{CEA}}
\and J.N.S.~Shapopi \inst{\ref{UNAM}}
\and A.~Sharma \inst{\ref{APC}}
\and H.~Sol \inst{\ref{LUTH}}
\and S.~Spencer \inst{\ref{ECAP}}
\and {\L.}~Stawarz \inst{\ref{UJK}}
\and R.~Steenkamp \inst{\ref{UNAM}}
\and S.~Steinmassl \inst{\ref{MPIK}}
\and C.~Steppa \inst{\ref{UP}}
\and H.~Suzuki \inst{\ref{Konan}}
\and T.~Takahashi \inst{\ref{KAVLI}}
\and T.~Tanaka \inst{\ref{Konan}}
\and A.M.~Taylor\protect\footnotemark[1] \inst{\ref{DESY}}
\and R.~Terrier \inst{\ref{APC}}
\and A.~Thakur \inst{\ref{Adelaide}}
\and M.~Tsirou \inst{\ref{DESY}}
\and C.~van~Eldik \inst{\ref{ECAP}}
\and M.~Vecchi \inst{\ref{Groningen}}
\and C.~Venter \inst{\ref{NWU}}
\and J.~Vink \inst{\ref{Amsterdam}}
\and H.J.~V\"olk \inst{\ref{MPIK}}
\and T.~Wach \inst{\ref{ECAP}}
\and S.J.~Wagner \inst{\ref{LSW}}
\and A.~Wierzcholska \inst{\ref{IFJPAN}}
\and M.~Zacharias \inst{\ref{LSW},\ref{NWU}}
\and A.A.~Zdziarski \inst{\ref{NCAC}}
\and A.~Zech \inst{\ref{LUTH}}
\and N.~\.Zywucka \inst{\ref{NWU}}
}

\institute{
Dublin Institute for Advanced Studies, 31 Fitzwilliam Place, Dublin 2, Ireland \label{DIAS} \and
Max-Planck-Institut f\"ur Kernphysik, Saupfercheckweg 1, 69117 Heidelberg, Germany \label{MPIK} \and
Yerevan State University,  1 Alek Manukyan St, Yerevan 0025, Armenia \label{Yerevan} \and
Landessternwarte, Universit\"at Heidelberg, K\"onigstuhl, D 69117 Heidelberg, Germany \label{LSW} \and
Kapteyn Astronomical Institute, University of Groningen, Landleven 12, 9747 AD Groningen, The Netherlands \label{Groningen} \and
Laboratoire Leprince-Ringuet, École Polytechnique, CNRS, Institut Polytechnique de Paris, F-91128 Palaiseau, France \label{LLR} \and
University of Namibia, Department of Physics, Private Bag 13301, Windhoek 10005, Namibia \label{UNAM} \and
Centre for Space Research, North-West University, Potchefstroom 2520, South Africa \label{NWU} \and
Deutsches Elektronen-Synchrotron DESY, Platanenallee 6, 15738 Zeuthen, Germany \label{DESY} \and
Institut f\"ur Physik und Astronomie, Universit\"at Potsdam,  Karl-Liebknecht-Strasse 24/25, D 14476 Potsdam, Germany \label{UP} \and
Université Paris Cité, CNRS, Astroparticule et Cosmologie, F-75013 Paris, France \label{APC} \and
Department of Physics and Electrical Engineering, Linnaeus University,  351 95 V\"axj\"o, Sweden \label{Linnaeus} \and
Institut f\"ur Physik, Humboldt-Universit\"at zu Berlin, Newtonstr. 15, D 12489 Berlin, Germany \label{HUB} \and
Laboratoire Univers et Théories, Observatoire de Paris, Université PSL, CNRS, Université Paris Cité, 5 Pl. Jules Janssen, 92190 Meudon, France \label{LUTH} \and
Sorbonne Universit\'e, CNRS/IN2P3, Laboratoire de Physique Nucl\'eaire et de Hautes Energies, LPNHE, 4 place Jussieu, 75005 Paris, France \label{LPNHE} \and
IRFU, CEA, Universit\'e Paris-Saclay, F-91191 Gif-sur-Yvette, France \label{CEA} \and
University of Oxford, Department of Physics, Denys Wilkinson Building, Keble Road, Oxford OX1 3RH, UK \label{Oxford} \and
Friedrich-Alexander-Universit\"at Erlangen-N\"urnberg, Erlangen Centre for Astroparticle Physics, Nikolaus-Fiebiger-Str. 2, 91058 Erlangen, Germany \label{ECAP} \and
Astronomical Observatory, The University of Warsaw, Al. Ujazdowskie 4, 00-478 Warsaw, Poland \label{UWarsaw} \and
Instytut Fizyki J\c{a}drowej PAN, ul. Radzikowskiego 152, 31-342 Krak{\'o}w, Poland \label{IFJPAN} \and
School of Physics, University of the Witwatersrand, 1 Jan Smuts Avenue, Braamfontein, Johannesburg, 2050 South Africa \label{Wits} \and
Laboratoire Univers et Particules de Montpellier, Universit\'e Montpellier, CNRS/IN2P3,  CC 72, Place Eug\`ene Bataillon, F-34095 Montpellier Cedex 5, France \label{LUPM} \and
School of Physical Sciences, University of Adelaide, Adelaide 5005, Australia \label{Adelaide} \and
Institut f\"ur Astronomie und Astrophysik, Universit\"at T\"ubingen, Sand 1, D 72076 T\"ubingen, Germany \label{IAAT} \and
Universit\"at Innsbruck, Institut f\"ur Astro- und Teilchenphysik, Technikerstraße 25, 6020 Innsbruck, Austria \label{Innsbruck} \and
Universit\"at Hamburg, Institut f\"ur Experimentalphysik, Luruper Chaussee 149, D 22761 Hamburg, Germany \label{UHH} \and
Obserwatorium Astronomiczne, Uniwersytet Jagiello{\'n}ski, ul. Orla 171, 30-244 Krak{\'o}w, Poland \label{UJK} \and
Universit\'e Bordeaux, CNRS, LP2I Bordeaux, UMR 5797, F-33170 Gradignan, France \label{CENBG} \and
Institute of Astronomy, Faculty of Physics, Astronomy and Informatics, Nicolaus Copernicus University,  Grudziadzka 5, 87-100 Torun, Poland \label{NCUT} \and
Aix Marseille Universit\'e, CNRS/IN2P3, CPPM, Marseille, France \label{CPPM} \and
Nicolaus Copernicus Astronomical Center, Polish Academy of Sciences, ul. Bartycka 18, 00-716 Warsaw, Poland \label{NCAC} \and
Department of Physics and Astronomy, The University of Leicester, University Road, Leicester, LE1 7RH, United Kingdom \label{Leicester} \and
Université Savoie Mont Blanc, CNRS, Laboratoire d'Annecy de Physique des Particules - IN2P3, 74000 Annecy, France \label{LAPP} \and
Yerevan Physics Institute, 2 Alikhanian Brothers St., 0036 Yerevan, Armenia \label{Yerevan2} \and
Department of Physics, Konan University, 8-9-1 Okamoto, Higashinada, Kobe, Hyogo 658-8501, Japan \label{Konan} \and
Kavli Institute for the Physics and Mathematics of the Universe (WPI), The University of Tokyo Institutes for Advanced Study (UTIAS), The University of Tokyo, 5-1-5 Kashiwa-no-Ha, Kashiwa, Chiba, 277-8583, Japan \label{KAVLI} \and
GRAPPA, Anton Pannekoek Institute for Astronomy, University of Amsterdam,  Science Park 904, 1098 XH Amsterdam, The Netherlands \label{Amsterdam}
}

\offprints{H.E.S.S.~collaboration,
\protect\\\email{\href{mailto:contact.hess@hess-experiment.eu}{contact.hess@hess-experiment.eu}};
\protect\\\protect\footnotemark[1] Corresponding authors
}

\authorrunning{H.E.S.S. collaboration}

\date{Received 23 October 2024 / Accepted 27 January 2025}

\abstract{The results of the first extragalactic gamma-ray survey by the High Energy Stereoscopic System (H.E.S.S.) are presented. 
The survey comprises 2720\ hours of very high-energy gamma-ray observations of the extragalactic sky, recorded with H.E.S.S. from 2004 up to the end of 2012. These data have been re-analysed using a common consistent set of up-to-date data calibration and analysis tools.
From this analysis, a list of 23 detected objects, predominantly blazars, was obtained. This catalogue was assessed in terms of the source class populations that it contains.  The level of source parameter bias for the blazar sources, probed by this observational dataset, was evaluated using Monte-Carlo simulations. Spectral results obtained with the H.E.S.S. data were compared with the \textit{Fermi}-LAT catalogues to present the full gamma-ray picture of the detected objects. Lastly, this unique dataset was used to assess the contribution of BL Lacertae objects and flat-spectrum radio quasars to the extragalactic gamma-ray background light at several hundreds of giga-electronvolts. These results are accompanied by the release of the high-level data to the astrophysical community.} 

\keywords{gamma rays: general -- gamma-rays: galaxies -- galaxies: active -- galaxies: jets -- BL Lacertae objects: general}

\maketitle
\newpage

\section{Introduction} \label{intro}

Since 2004, the High Energy Stereoscopic System (H.E.S.S.) has observed the southern hemisphere sky in the very high-energy domain (VHE, E$>$ 100 GeV). The H.E.S.S. collaboration has conducted a large programme of observations of the extragalactic sky. These observations have led to the discovery, as of June 2024\footnote{see \url{http://tevcat.uchicago.edu/}}, of tens of sources.

One of the key aims of H.E.S.S.' observational programme has been the study of particle acceleration and the energy loss processes at play in extragalactic sources. Due to its relatively small field of view (FoV), H.E.S.S. has scheduled pointed extragalactic observations towards either bright X-ray sources or bright \textit{Fermi}-LAT gamma-ray sources. Additionally, H.E.S.S. has carried out several target-of-opportunity observations. Such an observation strategy has led to the successful detection of emission from a range of extragalactic object classes, though primarily of the blazar type. This has allowed for the characterisation of the spectrum and temporal variability of these sources. Along with an improved understanding of blazars, H.E.S.S. has also observed and characterised other types of extragalactic sources such as galaxy clusters \citep{2012A&A...545A.103H,2012ApJ...750..123A,2009A&A...502..437A}, starburst galaxies \citep{2018A&A...617A..73H}, radio galaxies \citep[RGs;][]{2012ApJ...746..151A,2018A&A...619A..71H,2024A&A...685A..96H}, and gamma-ray bursts \citep{2009A&A...495..505A,2019Natur.575..464A,2021Sci...372.1081H}.

 The focus of this paper is to present the H.E.S.S. Extragalactic Survey (HEGS), a unique dataset taken with a range of different scientific purposes and observation strategies, analysed and presented collectively in one single coherent manner for the first time. All the data utilised in this study were taken with H.E.S.S. in its initial homogeneous configuration \citep{aha2006}, from 2004 to the end of 2012. The addition of a fifth, larger telescope, located at the centre of the array, demarcated the end of the homogeneous first phase of the instrument's lifetime, referred to as H.E.S.S.-I. All data taken during this H.E.S.S.-I phase are here re-analysed with the latest stable H.E.S.S. analysis software dedicated to this configuration to ensure homogeneous data treatment and consistent results.
 
The HEGS observational dataset offers a unique opportunity to use the full suite of H.E.S.S.-I extragalactic observations to probe different areas of modern astrophysics (emission mechanisms, variability, population of sources) and fundamental physics, such as the extragalactic background light \citep[EBL, e.g.][]{2007A&A...475L...9A,2013A&A...550A...4H,2017A&A...606A..59H,2024A&A...685A..96H}, the extragalactic gamma-ray background, intergalactic magnetic fields \citep[e.g.][]{2023ApJ...950L..16A}, a hypothetical violation of the Lorentz invariance \citep[e.g.][]{2011APh....34..738H,2019ApJ...870...93A}, the nature of dark matter \citep[e.g.][]{2018JCAP...11..037A,2018PhRvL.120t1101A,2021PhRvD.103j2002A,2022PhRvL.129k1101A}, or signs of new physics \citep[e.g. axion-like particles,][]{2013PhRvD..88j2003A,Cecil:2023vV}. Furthermore, with such a large dataset, it is possible for the first time to compare the H.E.S.S. extragalactic source population with the population probed at lower gamma-ray energies by \textit{Fermi}-LAT.

This paper is structured as follows. Section \ref{ObsMeth} introduces the full dataset used in this study, along with the analysis methods used. Section \ref{HEGScat} presents the HEGS catalogue. Section \ref{HEGSInterp} is devoted to the scientific interpretation of the results, and section \ref{summary} gives a summary of the results. This work is accompanied by the release of different VHE maps as well as spectral and temporal results of the detected objects. The released material is described in Appendix \ref{online}. 

\section{Observations and analysis}\label{ObsMeth}

\subsection{The High Energy Stereoscopic System}

H.E.S.S. is an array of Imaging Atmospheric Cherenkov Telescopes (IACTs) located in the Khomas highlands, in Namibia, at an altitude of 1800~m above sea level.
The H.E.S.S. telescopes detect the Cherenkov light that is emitted by charged particles created during the development of extensive air showers, produced by the interaction of high-energy particles (including gamma rays) with the high-altitude component of the atmosphere.

During the first phase of H.E.S.S. ~(H.E.S.S.-I, January 2004 - January 2013), the array was composed of four 12-m diameter telescopes, placed on a square with a side length of 120~m. In 2012, a bigger telescope, 28-m in diameter, was added to the centre of the array (the start of the H.E.S.S.-II phase of the instrument). The data presented here were accumulated during  H.E.S.S.-I.
The data taking procedure required the coincident detection of an atmospheric air-shower by at least two telescopes in order to trigger subsequent data acquisition. This requirement allows a stereoscopic reconstruction of the events \citep{1997APh.....8....1D}, providing considerably improved hadronic shower rejection power, better angular resolution, and energy reconstruction, compared to single-telescope observations \citep{2003ICRC....4.2323M,aha2006}.

\subsection{Data selection and clustering}

Each observation made by H.E.S.S. ~consists of data taken towards a given sky position during a nominal $28$ minute period, which defines a single observation run. Runs can be shorter for various reasons (end of night, clouds, technical issues), and those with a duration smaller than 5 minutes were excluded from this analysis.

In order to reduce systematic effects, only data taken under good atmospheric and instrumentation conditions were used \citep[see ][]{aha2006}. Large-zenith-angle observations, that is $>60^{\circ}$, were also discarded due to the large systematical uncertainties such observations can introduce \citep{2005A&A...437...95A, 2020A&A...635A.158M}.
The data selection was carried out by calibrating the remaining dataset with two independent calibration chains. Each calibration chain uses its own independent data selection criteria.
Only the runs selected as good quality by both calibration chains were retained for the analysis.

Contrary to the H.E.S.S. Galactic Plane Survey \citep[HGPS,][]{HGPS_paper}, the extragalactic observations were not performed in a systematic fashion to cover a pre-defined area of the sky. The bulk of the dataset consists of pointed observations, made towards promising target sources, Target-of-Opportunity observations, and deep monitoring of bright sources such as PKS~2155-304. As a consequence, the HEGS observations, are scattered throughout the region of the sky visible to H.E.S.S., leading to a highly non-uniform exposure distribution.

One of the first steps in the data analysis procedure was to group all the runs into observation clusters. This clustering was done using the {\tt DBSCAN} algorithm \citep{10.1145/3068335} from the {\tt scikit-learn} library \citep{scikit-learn}. 
A search radius of $4^{\circ}$ around each pointing position was applied, resulting in any two runs whose angular separation in their pointing positions was less than this being clustered together.

Following this clustering process, all clusters whose barycentre was located in the Galactic plane, that is the region of the sky bounded by $|b| < 10^{\circ}$ in Galactic coordinates, were discarded. Additionally, 
the Large and the Small Magellanic Clouds \citep{2015Sci...347..406H}, 47~Tucanae \citep{2009A&A...499..273A}, and SN~1006 \citep{2010A&A...516A..62A} regions, 
were also discarded from the dataset. We refer the reader to the dedicated analyses provided in the given references. Finally, clusters with fewer than 3 runs were discarded to avoid artefacts caused by potential calibration issues.

\begin{figure*}
\centering
\includegraphics[width=0.95\textwidth]{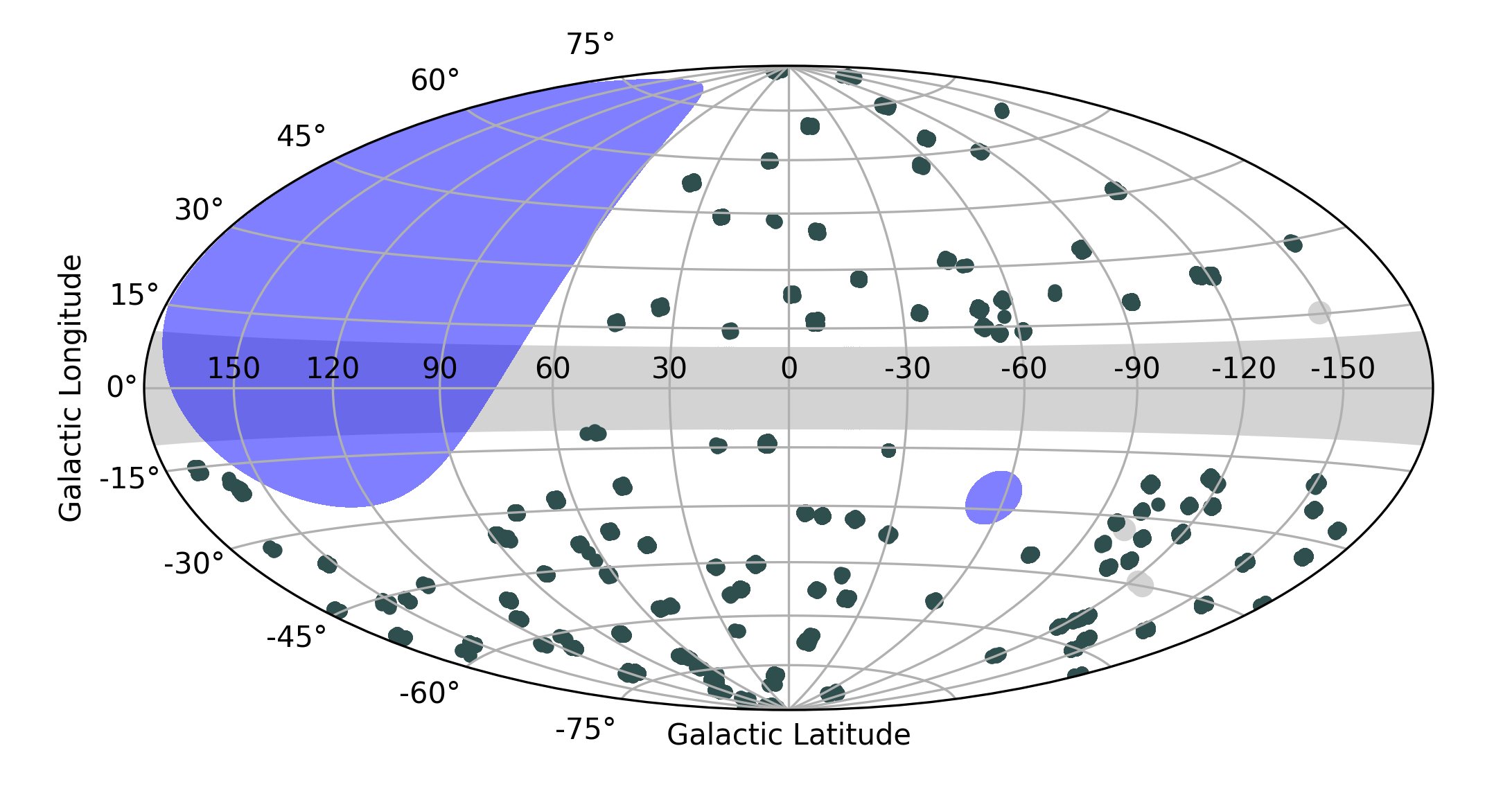}
\caption{Hammer-Aitoff sky map showing the HEGS clusters (dark green areas).
  The central grey band shows the regions with $|b| < 10^{\circ}$ that are discarded in this work. The other grey areas show the rest of discarded regions (see text). The blue areas denote the part of the sky not observable by H.E.S.S. , due to the zenith angles for observing these regions being above $ 60^{\circ}$.}
\label{fig:cluster_map}
\end{figure*}

As a result of the full clustering procedure, a total of 98 observational clusters was obtained. These cluster groups, each spatially separated from each other, were subsequently analysed independently.
Altogether, the cluster groups contain a total of 6500\ runs, and a corresponding total observation time of $\sim $2720\ hours. The collective sky coverage of these clusters covers approximately 5.7 \% of the total sky. The distribution of the clusters across the sky is shown in Fig.~\ref{fig:cluster_map} (dark green), together with the excluded sky regions (grey) and the inaccessible part of the sky (blue).

\subsection{Analysis and maps production}
\label{sec:maps}

The data were analysed using the likelihood reconstruction technique with Model++ analysis framework \citep{Naurois}, which provided the reconstruction of the events and the gamma-hadron separation. The set of cuts optimised for a low energy threshold, the so-called {\tt Loose} cuts \citep{aha2006}, were selected. 
The analysis procedure (maps, spectra, and light curves) was cross-checked with an independent calibration, reconstruction, and high-level analysis chain using the ImPACT method \citep{2014APh....56...26P}.

In this work, the sky map creation process for each cluster closely followed that adopted in the HGPS \citep[][section 3]{HGPS_paper}. For each observation, the events with a reconstructed angular distance larger than $2^{\circ}$ away from the pointing position were removed in order to avoid the introduction of systematic effects associated with observations at the edge of the fields of view. The global size of the maps for each cluster was defined to encompass all the selected events in that cluster.

The maps are divided into pixels of angular size $0.01^{\circ}\times0.01^{\circ}$, and each pixel in the maps represents a test region, centred on the pixel position, with a radius, $R_c = \sqrt{0.0125}~{\rm deg}$, corresponding to the value obtained as a result of the cuts optimisation for the detection of point-like sources in the {\tt Loose} cuts configuration of the Model analysis \citep{Naurois}.
This radius is designated as the oversampling radius. For each observation run, an energy threshold, defined as the energy at which the effective area is at $15\%$ of its maximum, was applied. As this increases with the angular distance to the pointing direction, it was evaluated at the largest value in this range ($2^{\circ}$), in order to be conservative. 

To assess the detection significance of gamma-ray hotspots in the skymap, the excess of gamma rays above background was estimated for each position in the map by using the ring background method \citep{2007A&A...466.1219B}. The ring used to estimate the background in the test region has an inner and an outer radius of $0.6^{\circ}$ and $1.2^{\circ}$, respectively. To avoid background estimation regions being contaminated by gamma rays, regions of $0.3^{\circ}$ radius around already known gamma-ray emitters were masked. The significance of the excess was then computed at each position using Eq. $17$ of \citet{Lima}.

\begin{figure*}
\centering
        \includegraphics[width=0.35\textwidth]{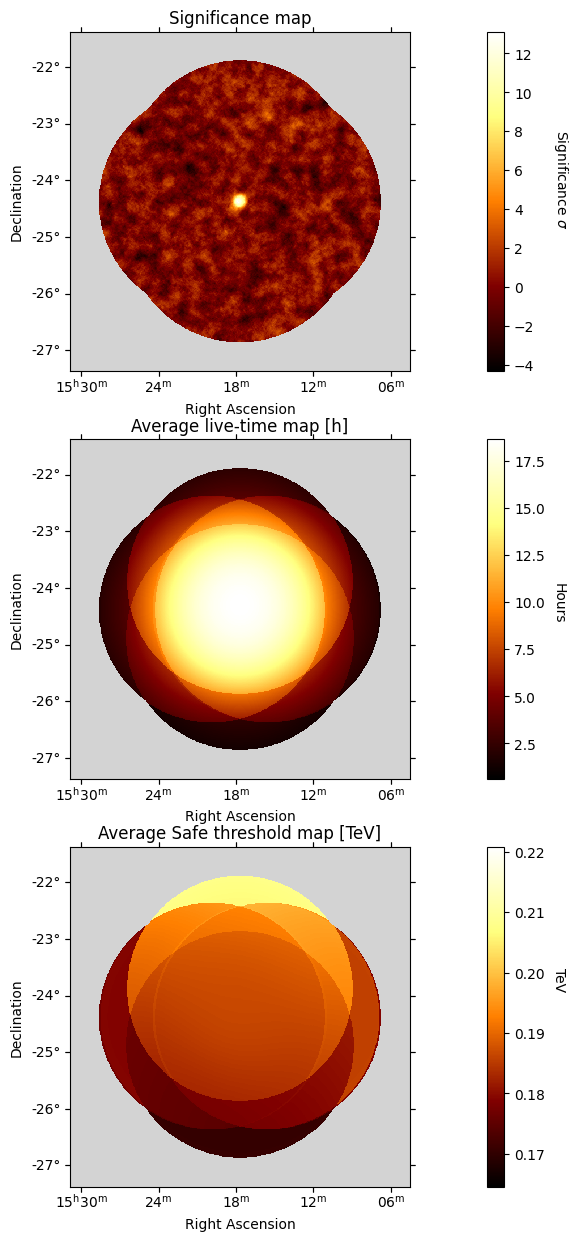}
        \includegraphics[width=0.45\textwidth]{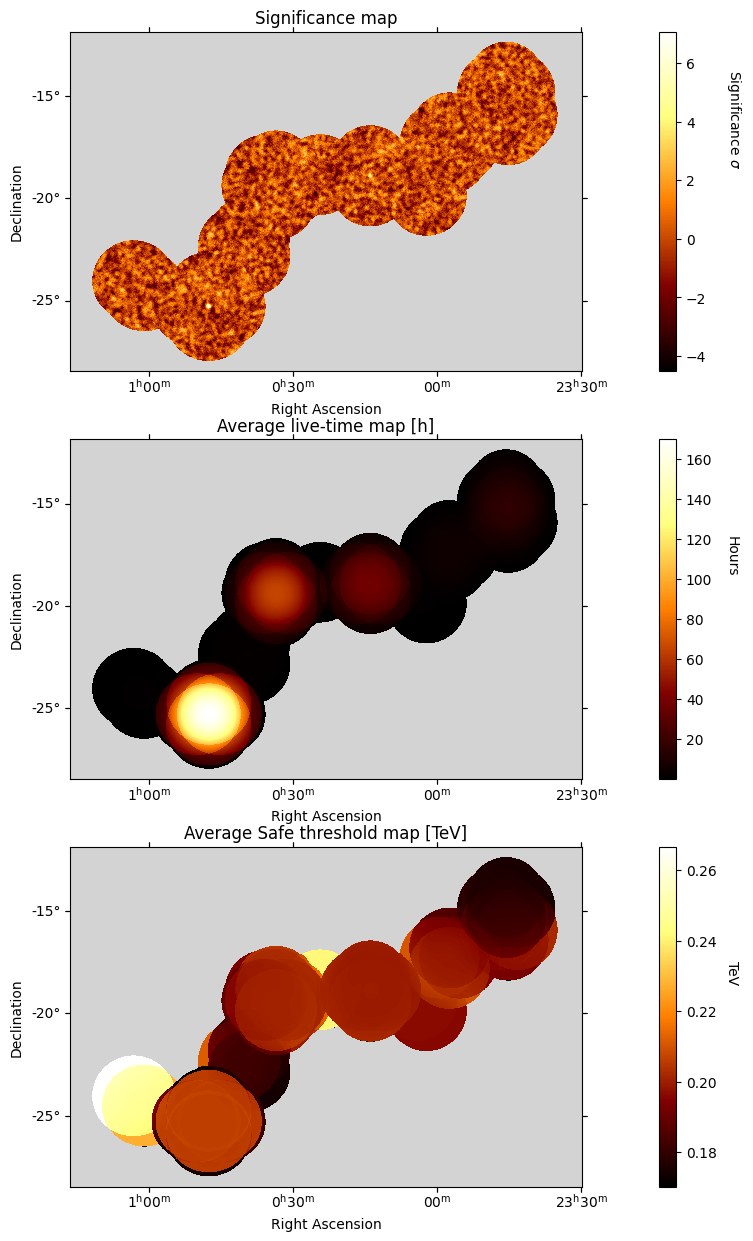}
    \caption{Examples of maps produced for this work. From top to bottom: Significance map, average live time map, and energy threshold map for two observation clusters (left and right panels). These maps are provided in the catalogue FITS files.}
    \label{fig:RC}
\end{figure*}

High-level maps (flux, flux error, upper limits, and sensitivity) were derived in the manner described in Sect. 3.4 of \citet{HGPS_paper}. For the calculation of these maps, a spectral index of $\Gamma = 3$ was chosen in order to be representative of the index of known extragalactic sources \citep{2013A&A...554A..75S}. The values shown in the maps are differential fluxes at an energy of $1~{\rm TeV}$. The upper limits shown are computed using \citet{2005NIMPA.551..493R}, and correspond to the 95\% confidence level (CL). The sensitivity represents the minimal flux that a source should have in order to reach a given statistical significance of detection above the background. The threshold for this analysis was set to the value of $5.7\sigma$, obtained as is described in Sect.~\ref{sub:sourcesignif}. An example of such high-level maps is presented in Fig.~\ref{fig:RC}.

\subsection{Significance calculation and source definition}
\label{sub:sourcesignif}

In order to determine the list of sources in the HEGS dataset, we determined the significance level over which the number of false detections is 1\% or less. This choice thus defines the significance threshold for source detection. 
In order to determine this threshold, the following Monte-Carlo approach was adopted. For each of the 98 HEGS clusters, a hundred simulated significance maps were computed using the same procedure as for the real HEGS data (see Sect.~\ref{sec:maps}). They are obtained using simulated counts maps drawn via Poisson sampling of the acceptance map (which represents the expected number of gamma-like background events in each pixel). From all the simulated fields, no more than one test position above $5.7 \sigma$ was found, and therefore we use this as the lower-limit criterion for source detection. In the following, it was therefore considered that a pixel contains a source candidate when its significance value was above $\sigma_{th} = 5.7$, with an estimated false detection probability\footnote{Given that there is only one occurrence of the $\sigma_{th}$ value, the false detection probability is ($1^{+2.89}_{-0.9}$)/100 with error bars at 90\% CL \citep[see table 40.3 of][]{PDGPaper} and the quoted value is the upper edge of this interval.} of less than 3.89/100 at 90\% CL.

A dedicated analysis of each source detected in the maps was subsequently  performed in order to derive its spectrum and light curve. For all except one hotspot, the identified positions of hotspots above the detection threshold had a known associated source within the oversampling radius defined above. The analysis was subsequently performed at the position of this counterpart, consistent with that identified in the dedicated publication, using the same analysis cuts as for the maps derivation. Significant excess, however, was also found for one hotspot above the detection threshold, whose location had no previously known VHE $\gamma$-ray emitter \footnote{It is found at position $\mathrm{RA} =$ \HMS{05}{52}{46.8}, $\mathrm{Dec} = $ \DMS{-32}{53}{42.0}.}. Since this hotspot was not found to be significant in the cross-check analysis, it was not considered further as a source candidate.

\subsection{Spectrum and light curve determination}
\label{sub:spec}

Spectral information for all the sources was extracted using the same analysis framework and the same sets of cuts as for the sky maps. The reflected background method \citep{2007A&A...466.1219B} was used to estimate the level of hadronic background contamination, and a forward-folding algorithm \citep{2001A&A...374..895P} was subsequently used to obtain the spectral model that best describes the VHE $\gamma$-ray~ emission and the corresponding best-fit parameters. Two different spectral models were used : a power-law model (PL) defined as $\phi=\phi_0 (E/E_0)^{-\Gamma}$, where $\phi_0$ is the normalisation at energy $E_0$, and $\Gamma$ is the spectral index; a log-parabola model (LP) defined as $\phi=\phi_0 (E/E_0)^{-\alpha - \beta \log(E/E_0)}$, where $\alpha$ describes the spectral index at energy $E_{0}$, and $\beta$ describes the curvature in the spectrum. A systematic uncertainty of 20\% on the flux and of 0.2 on the spectral index value are expected \citep{aha2006}.

To determine the preferred spectral model for each source, two nested models, which differ by one degree of freedom between them, were considered. A likelihood ratio test was applied using the test statistic (TS) difference between the two model fits. A value of 9 for this TS difference (corresponding\footnote{This estimator behaves as a $\chi^2$ distribution with one degree of freedom in the high-statistic regime.} to approximately $3\sigma$), has been used as a determining criterion for the preference of the LP model over the PL model.

The estimation of the flux points follows a similar procedure to that used in the HGPS paper: the energy range was divided into a set of energy bins, uniformly spaced in the logarithm of the energy, with each bin bounded by energies $E_i$, $E_{i+1}$. The differential flux in the energy bin was determined using the maximum-likelihood forward-folding algorithm at a reference energy, $E_{ref}=\sqrt{E_i\cdot E_{i+1}}$. Only the differential flux within the energy bin was left free to vary, with the spectral index being fixed to the best-fit value\footnote{For PKS~2155-304, the spectral model adopted was the log-parabola, with spectral parameters frozen to the best-fit values.}. An upper limit at the 95\% confidence level on the differential flux was computed for energy bins with a significance of less than 2$\sigma$. 

From the best-fit spectral model, light curve points were computed per night using the integrated fluxes above $300~{\rm GeV}$.  This value, commonly chosen in earlier H.E.S.S. publications, is above the energy threshold for all the variable sources studied here.

\subsection{Variability searches}
\label{meth:variability}

Using the ON-OFF test, as is described in \citet{2020APh...11802429B}, the flux variability of each of the clusters on two different timescales was looked for; namely, run-wise (typically 28 minutes) and night-wise. For each test region in a cluster’s map, the arrival time of all the events from that region were binned according to the tested timescales and the ON-OFF test algorithm was applied. For each test region, this test estimates the significance of the excess in a given time sample with respect to all the other samples, using \citet{Lima}. This method allows blind searches of variability to be performed over the whole field-of-view of a cluster. No detection of a new object or transient phenomenon was achieved with the HEGS data analysis. However, several known VHE sources were found to be variable with this method, as is discussed in Sect.~\ref{variability}.

\begin{figure}
\centering
\includegraphics[width=0.45\textwidth]{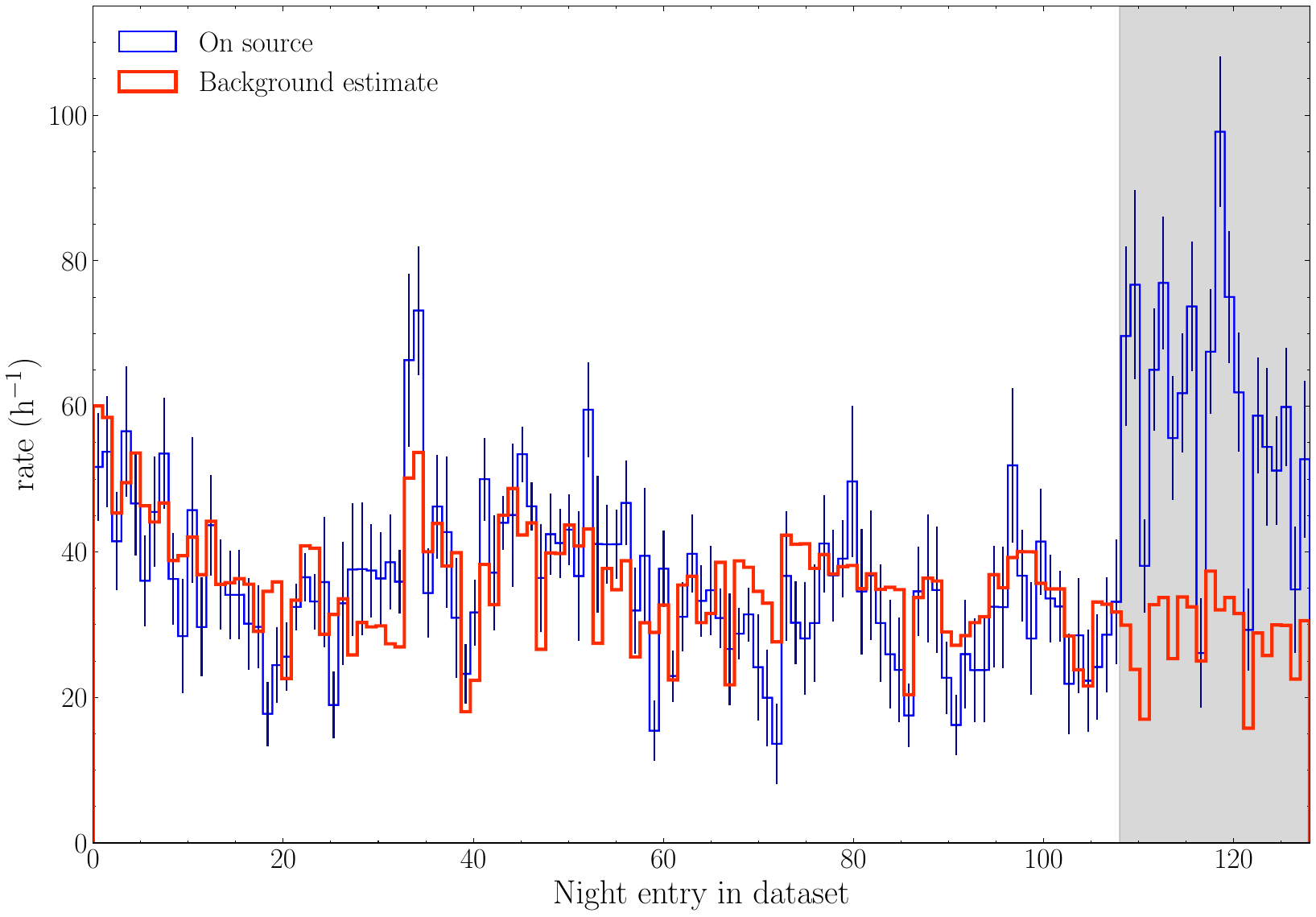}
\includegraphics[width=0.45\textwidth]{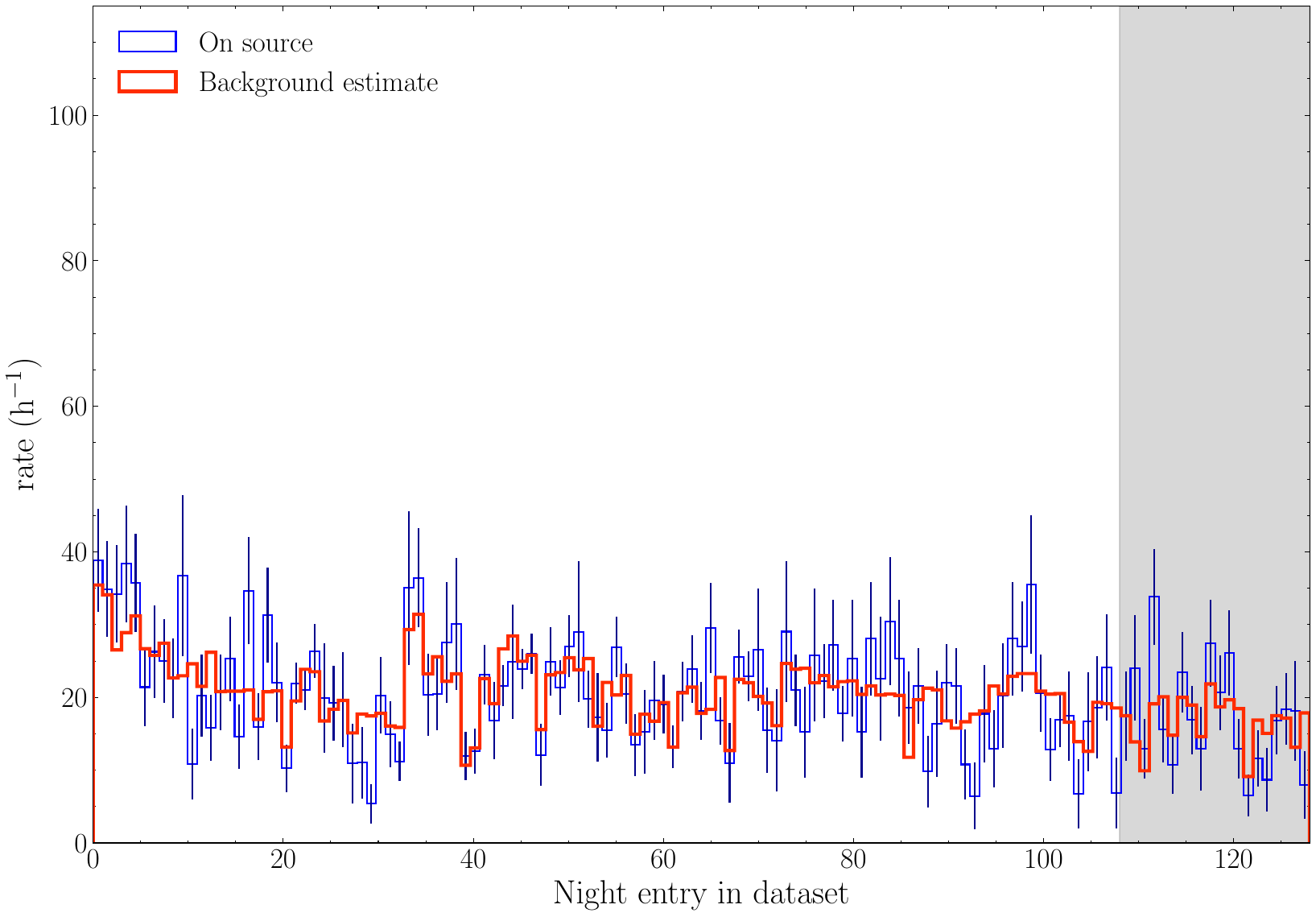}
\caption{Temporal distribution of the ON counts (blue) and the estimated background (red) for the observation of PKS~2005-489 (top panel) and a point 1 degree away (lower panel), as a function of the night number. The grey area is the time window when the source was significantly active.}
\label{fig:onoff_histo}
\end{figure}

A practical example showing the temporal structure of the flux from the source direction of the object PKS~2005-489 is shown in the top panel of Fig.~\ref{fig:onoff_histo}. In this figure the night-wise binning is adopted, with the above described variability search being performed, finding the source to be in a high flux state (indicated by the grey area in the figure) at the end of the observation period. This result is in agreement with the dedicated publication \citep{2011A&A...533A.110H}. The lower panel of this figure shows the result of the same analysis performed for a sky position 1~degree away from PKS~2005-489.
 
For the sources identified in the HEGS dataset, searches for variation in the flux across the full observation period were also performed for the night-wise light curve points by two additional methods: using the fractional variability, $\mathrm{F}_{\mathrm{var}}$, and its associated uncertainty, $\delta\mathrm{F}_{\mathrm{var}}$ \citep{Vaughan}; and fitting the flux with a constant and computing the corresponding $\chi^2$.
A source is estimated to be variable if either the ON-OFF test reaches $5\sigma$, the $\chi^2$ constant fit is rejected at more than $5\sigma$ or if the ratio $\mathrm{F}_{\mathrm{var}}/\delta\mathrm{F}_{\mathrm{var}}$ is above 5.

\subsection{Consistency checks}

Besides the re-running of the full analysis with an independent pipeline that was mentioned in Sect.~\ref{sec:maps}, several additional consistency checks were performed in order to verify the results of the maps derivation and spectral extraction procedure described above. Firstly, the distribution of significances from all the maps, outside the excluded regions, follows the expected normal distribution shape, with a mean of $-0.0348\pm0.0004$ and a width of $1.0267\pm0.0004$. 
These results are compatible with a level of background systematics of $\sim 2\%$, as was derived from Eq. $3$ of \citet{2021Univ....7..421D}.
Furthermore, the flux upper limit maps provide values compatible with those obtained in \citet{2014A&A...564A...9H}, in which the upper limits for 47 sources were derived. 

\begin{figure*}
\centering
        \includegraphics[width=0.95\textwidth]{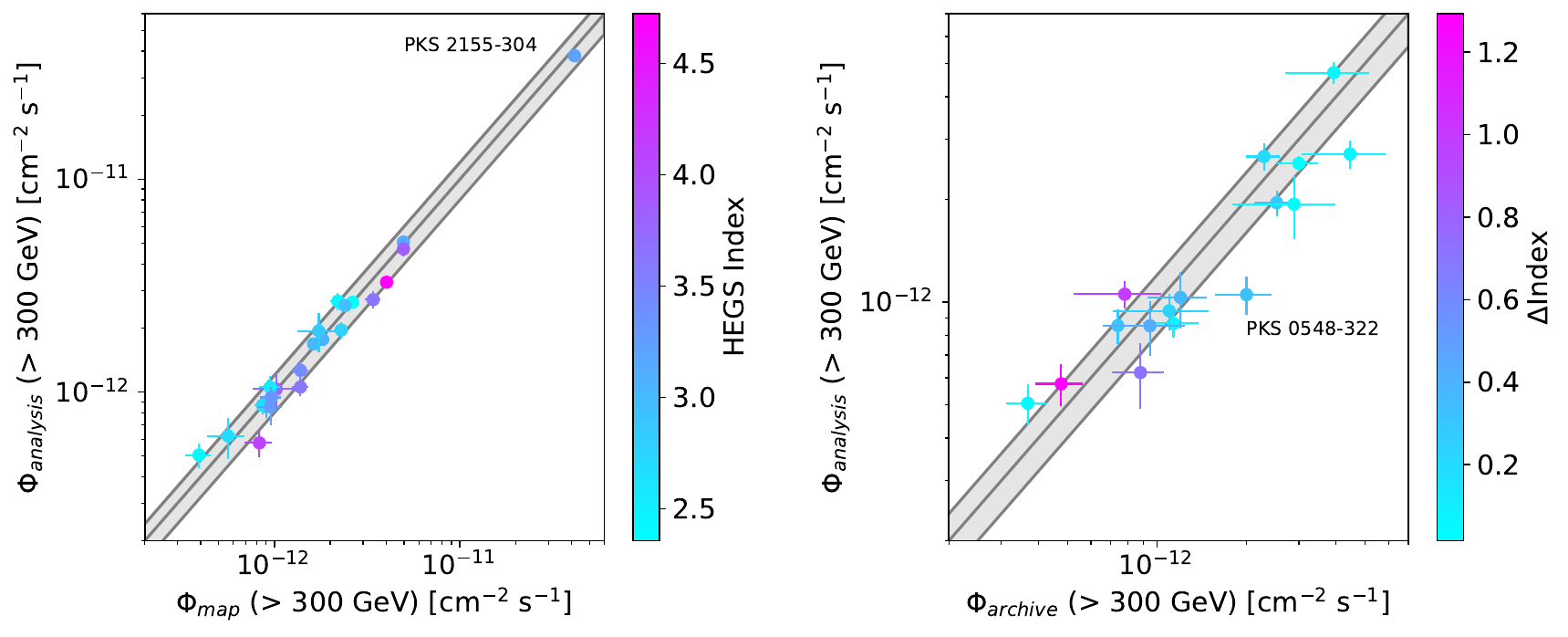}
    \caption{Comparison of the HEGS catalogue results with results from the HEGS maps and results from earlier publications. Left: Integral flux above 300 GeV obtained with the HEGS maps (with a spectral index fixed to 3) compared to the integral flux above 300 GeV determined in the dedicated analysis. The colour indicates the spectral index value obtained in the dedicated analysis. Right: Integral flux above 300 GeV obtained with the HEGS analysis compared to the integral flux above 300 GeV provided in earlier publications (see list in Table~\ref{table:list}), removing the sources found to be variable (see list in Table~\ref{table:variablesources}). The colour indicates the absolute difference in spectral index values from the HEGS catalogue and those from archival publications. The point on the right top corner of the left plot is PKS~2155-304 and has been marked for the reader as well as PKS 0548-322 on the right plot. The grey area represents the 20\% systematic uncertainties estimated by H.E.S.S. }
    \label{fig:pubflux}
\end{figure*}

Secondly, the flux values obtained in the flux maps for the detected sources are compatible with the values obtained in the subsequent dedicated analysis (Fig.~\ref{fig:pubflux}, left). Good agreement between the flux values from these analyses is found, with the residual difference below the $20\%$ systematic uncertainty level. These differences are mainly due to the assumption of a fixed spectral index made for the computation of the HEGS flux maps. When varying the assumed spectral index by $0.2$, the flux values in the maps on average varied by $\pm 15\%$.
The sky localisation of the signal detected for each source identified was fitted under the hypothesis of a point-like source. For all sources, the centroid of emission was compatible at the $68\%$ CL with the position of the identified counterpart, when considering the pointing systematics of $20"$ per axis from \citet{2010MNRAS.402.1877A}.
The angular extents of the detected sources were all compatible with the point-like hypothesis considering the performances of the analysis chain used for this study \citep{Stycz16}.

Lastly, the spectra of the sources from the HEGS analyses have been compared with those obtained in archival H.E.S.S. publications, as is shown in Fig.~\ref{fig:pubflux} (right). The sources included in this comparison are those listed in Table~\ref{table:list}, excluding those found to be variable (see Sect.~\ref{sec:var}). While the agreement is good, the remaining differences in the source fluxes seen in this figure can be attributed to the different calibration, analysis software, or energy threshold of HEGS with respect to those of the archival data.
One source in this comparison presents different spectral results at more than $1\sigma$ in HEGS with respect to the original publication: PKS~0548-322. The live-time reported in this work is about 3.5 times larger than in the previous publication with data taken until November 2008 \citep{2010A&A...521A..69A}.

In addition, for each source, the light curves derived from the dedicated spectral analysis are compatible with the flux values from the maps computed on the same timescales (run, night, moon-period).
Only in $0.2\%$ of the comparisons was the relative difference in the derived fluxes above the 1 sigma of statistical error level.

\section{The HEGS sources and their properties}\label{HEGScat}

\subsection{Source list}

The analysis of the H.E.S.S.-I observations yields the detection of 23 sources, all previously known as VHE emitters,  which compose the HEGS catalogue. 
Source characteristics (name, counterpart, position, redshift) are provided in Table~\ref{table:list}. The vast majority of sources (18 of the 23) are BL~Lac\ objects. Additionally, two RGs, one flat-spectrum radio quasar (FSRQ), one starburst galaxy, and one source of uncertain type complete the HEGS catalogue. A more detailed discussion of the source classes is made in Sect.~\ref{Sources}.

Almost all the extragalactic sources previously detected by H.E.S.S. are present in the HEGS catalogue with the notable exception of KUV~00311-1938. This high-frequency peaked BL~Lac object was found to be a tera-electronvolt emitter in \cite{2020MNRAS.494.5590A} at a level of 5.2 $\sigma$ with a soft spectrum. The publication reporting the detection includes observations taken in 2013 and 2014 which were not included in the HEGS observation dataset analysed (see Sect.~\ref{ObsMeth}). In the HEGS maps, the significance of KUV~00311-1938 is 1.6 $\sigma$

In Table~\ref{table:results}, the detection significance of the sources obtained from a dedicated analysis is provided. This value differs from the map results. This is expected since the dedicated analysis is performed at the position of the identified counterpart and not at the position of the maximum significance found in the maps. For 1ES~1312-423, due to the 2\mbox{$^\circ$} angular cuts for events selection, some runs taken at high offsets were only partially contributing events to the source region and were not included in the list of runs used for the dedicated analysis.

Each detected source has been assigned a source name in the format HESS JHHMM$\pm$DDd, where HHMM and $\pm$DDd are the source coordinates from the International Celestial Reference System (ICRS) in right ascension and declination, respectively. The naming was made based on the position of the source reported in the discovery papers (see Table~\ref{table:list}).

\begin{table*}
\small{
\caption{List of objects above the detection threshold.}
\label{table:list}
\begin{center}
\begin{tabular}{c c c c c c c }
\hline\hline 
H.E.S.S. name & Association & RA & Dec & Type & Redshift  & Reference \\\hline
HESS J0013-189 & SHBL J001355.9-185406  &  \HMS{00}{13}{56.04} &\DMS{-18}{54}{06.7}& BL Lac / HSP& 0.095 & 1 \\
HESS J0047-253& NGC 253  & \HMS{00}{47}{33.13} &\DMS{-25}{17}{19.7}& Starburst& 0.000864 & 2,3\\
HESS J0152+017& RGB J0152+017  & \HMS{01}{52}{39.61} &\DMS{+01}{47}{17.4}& BL Lac / HSP & 0.080 &4 \\
HESS J0232+202 & 1ES 0229+200  &  \HMS{02}{32}{48.61} &\DMS{+20}{17}{17.5}&BL Lac / HSP & 0.140 & 5\\
HESS J0303-241& PKS 0301-243 &  \HMS{03}{03}{26.50} &\DMS{-24}{07}{11.4}& BL Lac / HSP& 0.266 & 6\\
HESS J0349-119& 1ES 0347-121  & \HMS{03}{49}{23.19} &\DMS{-11}{59}{27.4}& BL Lac / HSP& 0.188 & 7\\
HESS J0416+010& 1ES 0414+009  & \HMS{04}{16}{52.49} &\DMS{+01}{05}{23.9}& BL Lac / HSP& 0.287 &  8 \\
HESS J0449-438& PKS 0447-439  & \HMS{04}{49}{24.70} &\DMS{-43}{50}{09.0}& BL Lac / HSP & 0.343 &  9 \\
HESS J0550-322& PKS 0548-322  &  \HMS{05}{50}{40.57} &\DMS{-32}{16}{16.5}&BL Lac / - & 0.069 &  10 \\
HESS J0627-354& PKS 0625-354  &      \HMS{06}{27}{06.73} &\DMS{-35}{29}{15.4}& RG& 0.055  &  11  \\
HESS J1010-313& 1RXS J101015.9-311909  &     \HMS{10}{10}{15.98} &\DMS{-31}{19}{08.4}& BL Lac / HSP& 0.143 &  12 \\
HESS J1103-234& 1ES 1101-232  &  \HMS{11}{03}{37.62} &\DMS{-23}{29}{31.2}& BL Lac / HSP& 0.186 &   13\\
HESS J1230+123& M 87   &  \HMS{12}{30}{49.42} &\DMS{+12}{23}{28.0}& RG &  0.0044 & 14,15  \\
HESS J1315-426& 1ES 1312-423 &  \HMS{13}{15}{03.39} &\DMS{-42}{36}{49.8}&BL Lac / HSP & 0.105 &16\\
HESS J1325-430& Centaurus A  & \HMS{13}{25}{27.62} &\DMS{-43}{01}{08.8}& RG & 0.00183 & 17,18 \\
HESS J1444-391& PKS 1440-389 & \HMS{14}{43}{57.20}	&\DMS{-39}{08}{40.1}&BL Lac / HSP & 0.139 &19\\
HESS J1512-090& PKS 1510-089 &        \HMS{15}{12}{50.53} &\DMS{-09}{05}{59.8}&FSRQ/ LSP & 0.361 & 20 \\
HESS J1517-243& AP Librae &     \HMS{15}{17}{41.81} &\DMS{-24}{22}{19.5}&BL Lac / ISP & 0.049 & 21\\
HESS J1555+111& PG 1553+113   &  \HMS{15}{55}{43.04} &\DMS{+11}{11}{24.4}& BL Lac / HSP& 0.433  & 22,23\\
HESS J2009-488& PKS 2005-489  &  \HMS{20}{09}{25.39} &\DMS{-48}{49}{53.7}& BL Lac / HSP& 0.071 & 24,25 \\
HESS J2158-302& PKS 2155-304   &     \HMS{21}{58}{52.07} &\DMS{-30}{13}{32.1}&BL Lac / HSP & 0.117  & 26,27\\
HESS J2324-406& 1ES 2322-409  & \HMS{23}{24}{44.67} &\DMS{-40}{40}{49.4}&BL Lac / HSP & 0.174  & 28 \\
HESS J2359-306& H 2356-309  &  \HMS{23}{59}{07.90} &\DMS{-30}{37}{40.7}&BL Lac / HSP & 0.165 & 29,30 \\
\hline

\end{tabular}
\end{center}
}

\tablefoot{
In addition to the H.E.S.S. name, the columns give the associated source name, its coordinates from the SIMBAD server \citep{SIMBAD}, type / SED class for the \textit{Fermi} catalogues, redshift and a discovery or reference paper by the H.E.S.S. collaboration.

 The redshifts of PG~1553+113 and PKS~1440-389 not known at the time of the discovery, were extracted from \citet{2022MNRAS.509.4330D} for PG~1553+113 and \citet{2021A&A...650A.106G} for PKS~1440-389.
}
\tablebib{
(1)~\citet{2013A&A...554A..72H}; (2) \citet{2009Sci...326.1080A}, (3) \citet{2012ApJ...757..158A}; (4) \citet{2008A&A...481L.103A};
(5) \citet{2007A&A...475L...9A}; (6) \citet{2013A&A...559A.136H}; (7) \citet{2007A&A...473L..25A}; (8) \citet{2012A&A...538A.103H};
(9) \citet{2013A&A...552A.118H}; (10) \citet{2010A&A...521A..69A}; (11) \citet{2018MNRAS.476.4187H}; (12) \citet{2012A&A...542A..94H};
(13) \citet{2007A&A...470..475A}; (14) \citet{2006Sci...314.1424A}; (15) \citet{2023A&A...675A.138H}; (16) \citet{2013MNRAS.434.1889H};
(17) \citet{2009ApJ...695L..40A}; (18) \citet{2018A&A...619A..71H}; (19) \citet{2020MNRAS.494.5590A}; (20) \citet{2013A&A...554A.107H};
(21) \citet{2015A&A...573A..31H}; (22) \citet{2006A&A...448L..19A}; (23)  \citet{2015ApJ...802...65A}; (24) \citet{2005A&A...436L..17A};
(25) \citet{2011A&A...533A.110H}; (26) \citet{1999APh....11..145C}; (27) \citet{2017A&A...598A..39H}; (28) \citet{2019MNRAS.482.3011A};
(29) \citet{2006A&A...455..461A}; (30) \citet{2010A&A...516A..56H}.
}
\end{table*}

\begin{table*}
\small{
\caption{Spectral results of sources detected in the HEGS FoV.}

\label{table:results}
\begin{center}
\begin{tabular}{c c c c c c c c c}
\hline\hline 
Name & $\sigma$ &$\phi_{\rm dec}(E_{\rm dec})$  & $\Gamma$ & $E_{\rm dec}$ &  $E_{\rm th}$ &$\phi_{\rm dec}(E_{\rm dec})$ EBL-corrected &  $\Gamma$ EBL-corrected  \\
 &  &  $10^{-12}\,{\rm cm^{-2}}\,{\rm s^{-1}}\,{\rm TeV^{-1}}$  &  & TeV & TeV & $10^{-12}\,{\rm cm^{-2}}\,{\rm s^{-1}}\,{\rm TeV^{-1}}$ & \\\hline

NGC 253 & 6.7 & 0.43 $\pm$ 0.06 & 2.43 $\pm$ 0.16 & 0.61 & 0.13 & 0.44   $\pm$0.06 & 2.34 $\pm$ 0.17   \\
RGB J0152+017 & 11.0 & 7.26 $\pm$ 0.71 & 3.42 $\pm$ 0.23 & 0.33  &0.20&9.47   $\pm$ 0.93& 2.92 $\pm$ 0.24   \\
1ES 0229+200 & 14.3 & 0.79 $\pm$ 0.07 & 2.78 $\pm$ 0.15 & 0.79  &0.32 &2.96   $\pm$0.26 & 1.74 $\pm$ 0.16   \\
PKS 0301-243 & 12.5 & 18.17 $\pm$ 1.7 & 3.6 $\pm$ 0.22 & 0.25  &0.13 &35.38   $\pm$ 3.35 & 2.28 $\pm$ 0.28   \\
1ES 0347-121 & 14.5 & 9.74 $\pm$ 0.68 & 3.07 $\pm$ 0.13 & 0.32  &0.16&19.65   $\pm$ 1.36 & 1.89 $\pm$ 0.15   \\
1ES 0414+009 & 10.0 & 3.48 $\pm$ 0.41 & 3.14 $\pm$ 0.24 & 0.36  &0.18& 13.23   $\pm$ 1.54  & 1.21 $\pm$ 0.28   \\
PKS 0447-439 & 20.5 & 33.6 $\pm$ 2.6 & 3.83 $\pm$ 0.20 & 0.32  &0.18 & 130.44   $\pm$ 10.01&  1.40 $\pm$ 0.22   \\
PKS 0548-322 & 5.4& 1.43 $\pm$ 0.18 & 2.54 $\pm$ 0.18 & 0.51  &0.16 &2.21   $\pm$ 0.29& 2.04 $\pm$ 0.18   \\
PKS 0625-354 & 5.7& 4.82 $\pm$ 1.01 & 2.86 $\pm$ 0.36 & 0.41  & 0.18 &6.32   $\pm$  1.32 & 2.47 $\pm$ 0.37   \\
1RXS J101015.9-311909 & 8.0& 7.93 $\pm$ 0.99 & 3.31 $\pm$ 0.26  & 0.29 & 0.15 &12.00   $\pm$  1.51&  2.43 $\pm$ 0.28   \\
1ES 1101-232 & 19.2& 9.67 $\pm$ 0.50 & 2.92 $\pm$ 0.08 & 0.36  &0.12 &21.99   $\pm$ 1.14 & 1.75 $\pm$ 0.10   \\
M 87 &18.5& 1.22 $\pm$ 0.07 & 2.36 $\pm$ 0.07 & 0.79 & 0.18&1.36  $\pm$  0.08  &2.25 $\pm$ 0.07   \\
1ES 1312-423 &  4.7& 3.51 $\pm$ 0.64 & 3.28 $\pm$ 0.39 & 0.36 & 0.15 &5.19  $\pm$  0.94 & 2.58 $\pm$ 0.41   \\
Centaurus A & 11.6 & 1.37 $\pm$ 0.12 & 2.67 $\pm$ 0.13 & 0.48 & 0.15 &1.12   $\pm$ 0.10& 2.58 $\pm$ 0.13   \\
PKS 1440-389 &12.3 & 28.5 $\pm$ 2.7 & 3.63 $\pm$ 0.24 & 0.29 & 0.16 &42.16  $\pm$  4.01& 2.86 $\pm$ 0.26   \\
PKS 1510-089 & 9.2& 22.6 $\pm$ 3.2 & 4.11 $\pm$ 0.44 & 0.22 & 0.15 &48.79  $\pm$ 6.85 & 2.16 $\pm$ 0.50   \\
AP Librae & 10.9& 3.85 $\pm$ 0.35 & 2.45 $\pm$ 0.11 & 0.49 & 0.13 &   5.15   $\pm$ 0.47& 2.09 $\pm$ 0.12   \\
PG 1553+113 &28.8 & 46.1 $\pm$ 2.5 & 4.72 $\pm$ 0.20 & 0.29 & 0.22 &304.01   $\pm$ 16.11& 1.49 $\pm$ 0.24   \\
PKS 2005-489 & 44.4 & 9.28 $\pm$ 0.27 & 3.13 $\pm$ 0.05 & 0.46  &  0.18  &13.73   $\pm$ 0.39&2.62 $\pm$ 0.06   \\
PKS 2155-304 & 516.2 & 336.2 $\pm$ 1.5 & 2.90 $\pm$ 0.01 & 0.29   & 0.12 &320.36  $\pm$   1.20& 2.51 $\pm$ 0.01   \\
 &  &  & 0.41 $\pm$ 0.01 &  &  & &   \\
1ES 2322-40.9 &6.8 & 13.0 $\pm$ 2.5 & 3.77 $\pm$ 0.51 & 0.28 & 0.15& 20.39  $\pm$ 3.85 & 2.80 $\pm$ 0.56   \\
H 2356-309 & 22.2& 8.96 $\pm$ 0.40 & 2.99 $\pm$ 0.08 & 0.32 & 0.12&16.68   $\pm$  0.74& 1.97 $\pm$ 0.09   \\
\hline

\end{tabular}
\end{center}
}
\tablefoot{ The columns give the names of the sources, the significance from the dedicated analysis, the flux at the decorrelation energy, the spectral index, the decorrelation energy in tera-electronvolts, and the energy threshold. The last columns give the flux at the decorrelation energy and the spectral index, both corrected for EBL absorption using the model of \citet{2011MNRAS.410.2556D}. Errors are statistical only. For PKS~2155-304, the first line gives the value of $\alpha$ and the second $\beta$ of the Log-parabola model and the last two columns are the EBL-corrected normalisation and index of the power-law fit.}
\end{table*}

\begin{figure*}
\centering
        \includegraphics[width=0.99\textwidth]{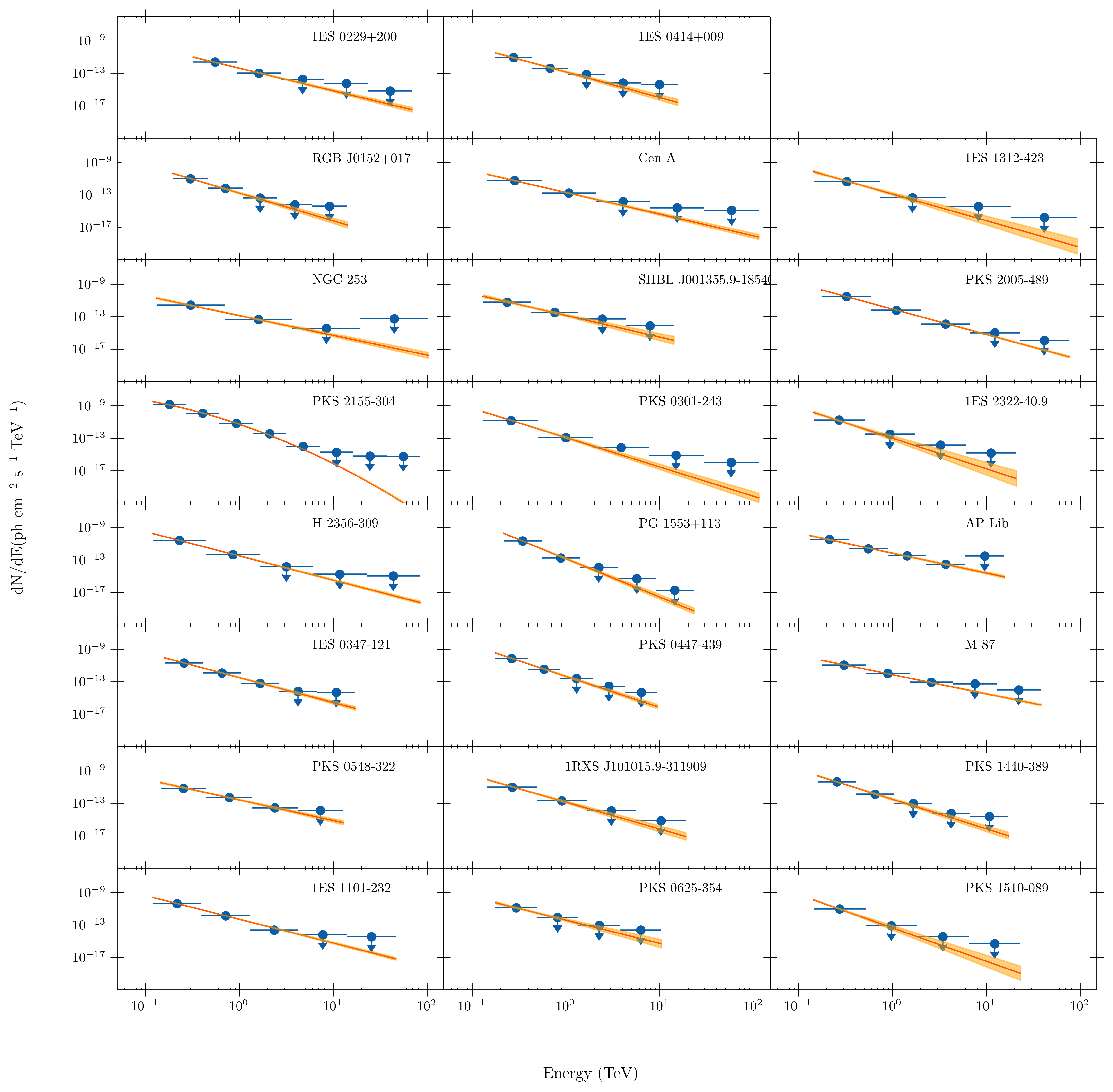}
    \caption{Spectra of the detected sources. The red line is the best fit model and the orange bands show the 1$\sigma$ contours of the likelihood fit. No EBL correction has been applied. The blue points are derived as explained in the text, and the arrows represent the 95\% CL upper limits.}
    \label{fig:HEGSspec}
\end{figure*}

The best fit spectral parameters for all detected sources are summarised in Table~\ref{table:results}. A power-law model was preferred for all sources, with the exception of PKS~2155-304 for which the log parabola model was preferred. For the BL~Lac\ sources a mean spectral index value of 3.25$\pm$ 0.53  was found. The spectral results are presented in Fig.~\ref{fig:HEGSspec}.

These spectral results are affected by EBL absorption. During their propagation over cosmological distances, VHE $\gamma$-rays\ interact with lower energy (optical and infrared) photons from the EBL. Such interactions lead to an energy- and redshift-dependent absorption of the gamma-ray emission. The modification of the spectra is taken into account by fitting an EBL absorbed PL spectral model: $ \Phi_{\rm E} =  \phi_0 (E/E_0)^{-\Gamma} \exp(-\tau(z))$. We note that the determination of $\tau$ suffers from uncertainties, following uncertainties in the EBL model. The prediction of current EBL models varies within a factor of around 10\% in the frequency range that attenuates VHE photons. Here we adopt the EBL model from \citet{2011MNRAS.410.2556D}. The results obtained are provided in Table~\ref{table:results}.

\subsection{Spectral properties of HEGS source classes in the Fermi 4FGL catalogue}\label{comparCat}

\begin{figure*}
\centering
        \includegraphics[width=0.49\textwidth]{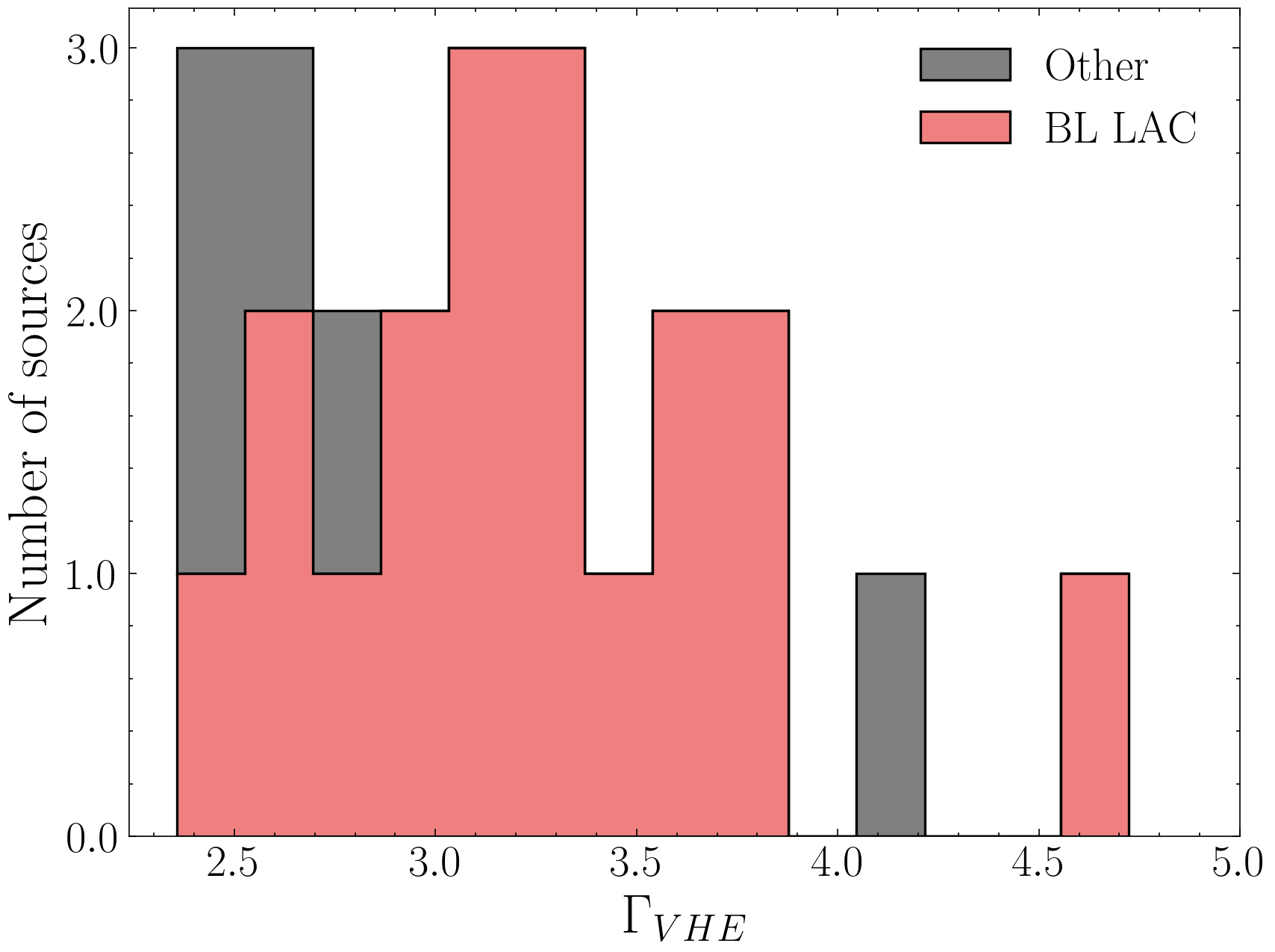}
        \includegraphics[width=0.49\textwidth]{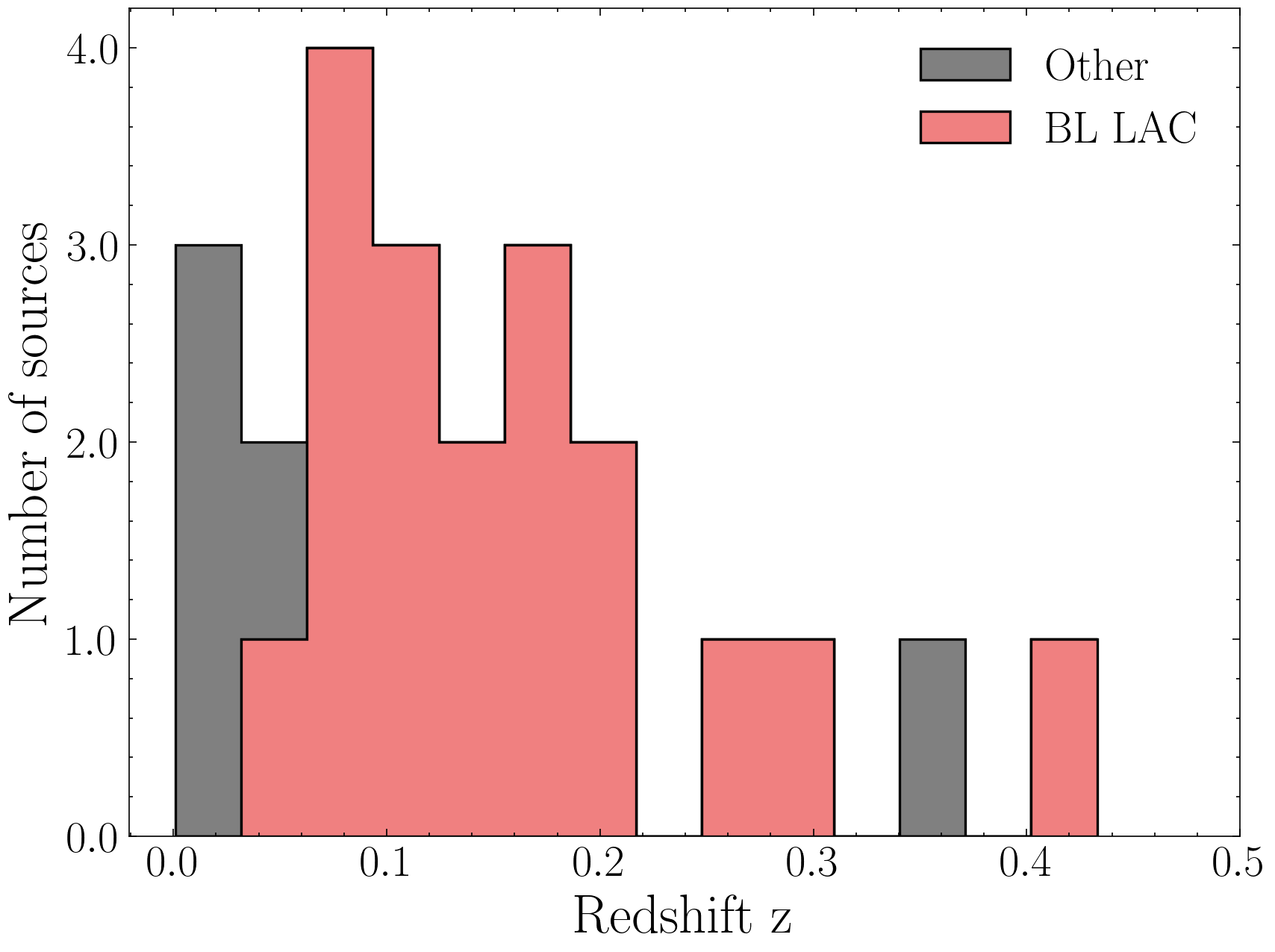}
    \caption{Distribution of photon index and redshift. Left: Distribution of the photon index, $\Gamma_{\rm VHE}$, measured in this study. Right: histogram of the redshift $z$ of the detected sources and presented in this work. The red bars are for the BL~Lac objects while the grey ones, on top of the former, give the rest of the sources.}
    \label{fig:HistoGammavhe}
\end{figure*}

The fourth \textit{Fermi}-LAT  catalogue of sources \citep[4FGL, ][]{2019arXiv190210045T} provides spectral information at lower energies (0.05~GeV--1~TeV) than those probed by the HEGS catalogue.  Table~\ref{table:MeanIndex} gives both the 4FGL observed and EBL absorption-corrected mean spectral index for the classes of sources detected in the HEGS catalogue. Fig.~\ref{fig:HistoGammavhe} shows the EBL absorption-corrected spectral indices distribution (left) and the redshift distribution (right). Due to the dominance of the BL~Lac objects in the HEGS catalogue, our conclusions, presented in the following, are predominantly drawn for this object class.

\begin{table*}
\small{
\caption{Mean power-law spectral indices in HE, VHE and VHE corrected for EBL domains.}
\label{table:MeanIndex}
\begin{center}
\begin{tabular}{c c c c c c}
\hline\hline 
Type & HE Mean& VHE Mean & VHE EBL-corrected Mean & z& Number of source\\\hline
BL~Lac~ & 1.83 $\pm$ 0.16 & 3.25$\pm$ 0.53 & 2.16$\pm$ 0.47 & [0.08-0.433] &   18 \\
Starburst& 2.14 $\pm$ 0.05  & 2.43$\pm$ 0.16 & 2.34$\pm$0.17 &  0.000864 &  1 \\
FSRQ& 2.37 $\pm$ 0.01 & 4.11 $\pm$  0.44  &  2.16 $\pm$  0.50 &  0.361  &  1 \\
RG& 2.20 $\pm$ 0.35 & 2.62 $\pm$ 0.59 & 2.43 $\pm$ 0.13 & [0.00183-0.05486] &  3 $\dagger$ \\ 
\hline 

\end{tabular}
\end{center}
}
\tablefoot{First column is the mean power-law spectral indices in HE (taken from the 4FGL), second and third columns are VHE and VHE corrected for EBL. Value are given for each class of blazars present in the HEGS catalogue. The standard deviation of the distributions are reported. For the starburst and FSRQ, we report the statistical error. The range of redshifts, $z$, is also given and the number of sources in the sample are given in the last two columns.
$\dagger$ PKS~0625-35 is included in this list.
}
\end{table*}

\begin{figure}
\centering
        \includegraphics[width=0.45\textwidth]{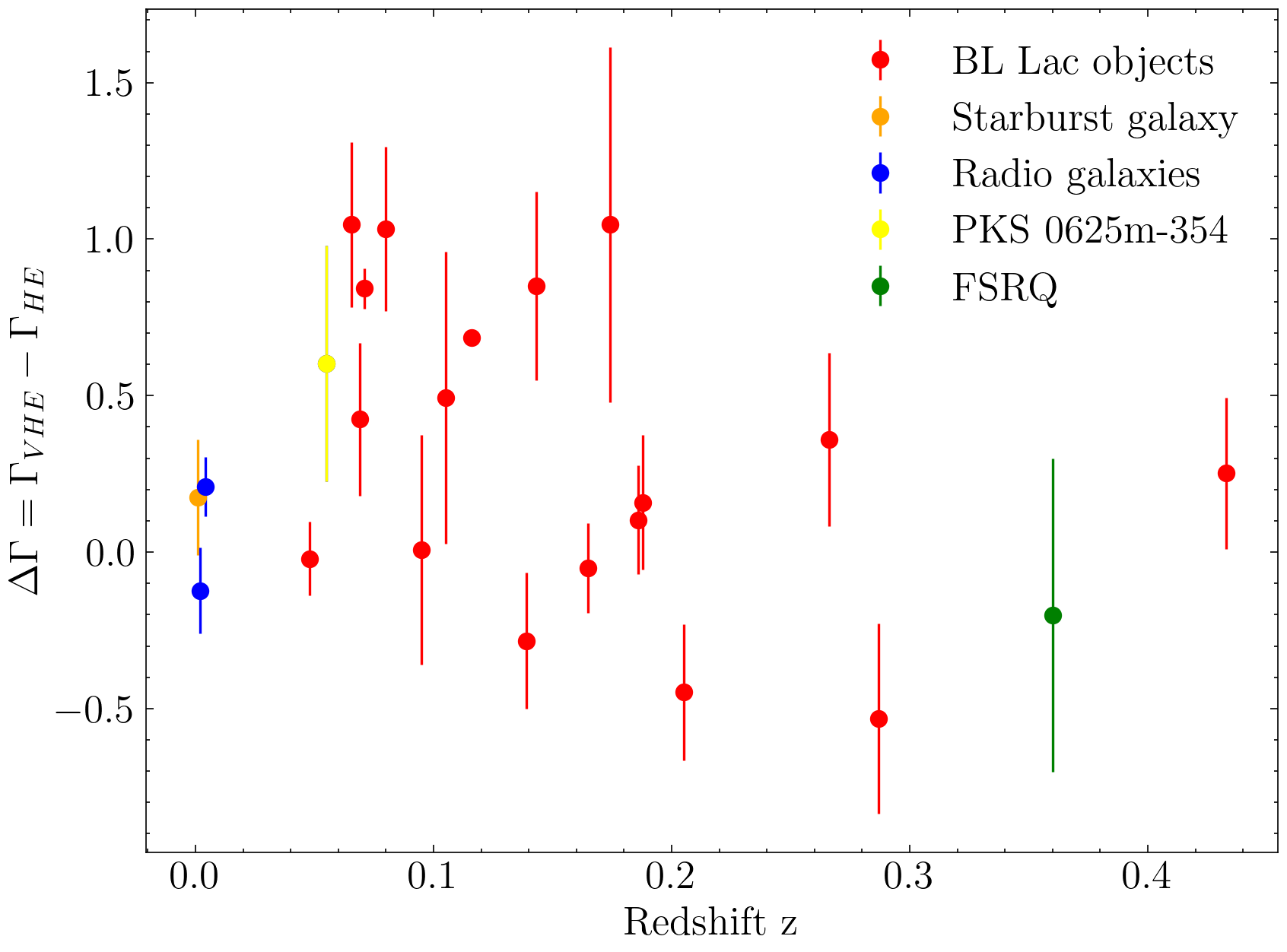}
    \caption{Difference between the VHE spectral index, corrected for EBL attenuation, and the spectral index reported in the 4FGL, $\Delta \Gamma$, as a function of the source redshift $z$}
    \label{fig:deltagamma}
\end{figure}

The EBL effects are known to be responsible for imprinting a trend between the redshift and the HE-VHE spectral break $\Delta \Gamma$ \citep[see e.g. ][]{2013A&A...554A..75S}. Nevertheless, once these EBL effects have been corrected for, if a remaining correlation between the two observables is found, this could potentially provide evidence for astrophysical effects such as redshift-dependent source evolution, or for the onset of new physics such as the existence of Axion-like particles \citep{2013PhRvD..87c5027M}. Fig.~\ref{fig:deltagamma} illustrates the search for such residual spectral trends with redshift, using the spectral results from the HEGS catalogue. (see Table~\ref{table:results} and 4FGL data). The VHE data in the HEGS catalogue appears consistent with no residual correlation between the break value $\Delta \Gamma$ and the source redshift.

Once EBL absorption is corrected for, the difference between the High Energy (HE, 100 MeV$<$E$<$100 GeV), and VHE intrinsic spectral shape of BL~Lac objects can be assessed. Assuming a PL spectral model in these two energy ranges, the difference between the VHE and HE spectral index, $\Delta \Gamma_{{\rm BL LAC}}$, can be used to compare the spectral differences in these energy bands. A mean value of $\Delta \Gamma_{{\rm BL~LAC}} = 0.33$ with a variance of $0.24$ is found from such a comparison. In the $\nu F_{\nu}$ representation, only BL~Lac objects have their peaks in the HE-VHE range. Deeper insight into this result, comparing H.E.S.S. results for different subclasses of BL~Lac, is currently inconclusive due to the low  statistics of these subclass populations. We therefore leave a study of this to future VHE catalogue studies, expected to achieve the larger statistics necessary.

\subsection{HE and VHE combined spectral fits}

With the aim of constraining the spectral energy distribution from the \textit{Fermi}-LAT  energy range to the H.E.S.S. maximal energy, the fourth data release (DR4) of the 4FGL \citep{2023arXiv230712546B} data points were combined with the H.E.S.S. data and fitted with a LP model including the effect of EBL absorption. Using this spectral model (with three free parameters), the parameter space was explored. The likelihood $\mathcal{L}_{H.E.S.S.}$ is the same as that minimised for the spectral parameters determination \citep{2001A&A...374..895P}. For the \textit{Fermi}-LAT  points, the likelihood log$(\mathcal{L}_{LAT})$ is 2 times the $\chi^2$ distance between the model and the \textit{Fermi}-LAT points.

The extrema of log$(\mathcal{L}_{H.E.S.S.})+$log$(\mathcal{L}_{LAT})$ for each source was obtained using the {\tt emcee} python package \citep{2013PASP..125..306F}. The results are presented in Table~\ref{table:sed}. Only the sources that are not found to be variable in the VHE range have been considered in the calculation since the HEGS and \textit{Fermi}-LAT data are not contemporaneous.  Centaurus A has also been excluded from this study since the emissions measured by \textit{Fermi}-LAT and H.E.S.S. have likely a different origin \citep{2020Natur.582..356H}. In total, 15 sources have been considered in this analysis.
For all the sources except three, namely SHBL~J001355.9-185406, 1ES~0229+200, and 1ES~0414+009, the value of $\beta$ is found to be significantly different from 0, considering only the statistical uncertainties. Physically, this means that the location of the peak position, in an spectral energy distribution (SED) representation, can be constrained. The constrained values are provided in Table~\ref{table:sed}.

The peak position in the SED for the majority of sources is found to lie in the range 10-100~GeV. This is consistent with the fact that for most of these sources the intrinsic spectral index in the 4FGL is $<2$, whereas in HEGS it is $>2$. The notable exception is Ap~Librae, the only low-frequency-peaked BL~Lac in HEGS, for which a peak position of $0.36^{+0.17}_{-0.11}$ GeV has been obtained. This is compatible with what has been found by \citet{2015A&A...573A..31H}. In the case of NGC~253 and PKS~1510-089, the peak SED position is at a lower energy than the \textit{Fermi}-LAT range and cannot be determined. For the case of 1ES~0414+009, the best fit model has a marginally negative $\beta$ value, preventing the peak value to be determined.

For the other sources (SHBL~J001355.9-185406, 1ES~0229+200, and 1ES 1101-232), a lower limit is given at a 99\% confidence level. For 1ES~0229+200 and 1ES~1101-232, \citet{2017arXiv171106282C} already found that the position of the peak energy flux is not constrained, although our limit value obtained here is more robust.

\begin{table*}
\small{
\caption{\textit{Fermi}-LAT  and H.E.S.S. combined fit results. }
\label{table:sed}
\begin{center}
\begin{tabular}{c c c c c c}
\hline\hline 
Name  &$\phi_{\rm 0}(E_{\rm 0})$  & $\alpha$ & $\beta$ & $E_{\rm 0}$ & $E_{\rm peak}$ \\
 &    $10^{-12}\,{\rm cm^{-2}}\,{\rm s^{-1}}\,{\rm TeV^{-1}}$   & &  & TeV  & GeV \\\hline

SHBL J001355.9-185406 & 2.24$\pm$ 0.63 &  2.05 $\pm$ 0.14 & 0.04  $\pm$ 0.05 & 0.362 & $<0.3\cdot 10^{3}$ \\
NGC 253 & 0.96$\pm$ 0.06  &  2.38$\pm$ 0.06 &  0.04 $\pm$ 0.02 & 0.426 & ---  \\
1ES 0229+200 & 1.63$\pm$ 0.03  & 1.76 $\pm$ 0.07 &  0.03 $\pm$ 0.04 & 0.607 & $<5.11\cdot 10^{3}$  \\
PKS 0301-243 & 37. $\pm$ 3.3  &  2.43$\pm$ 0.04&  0.14 $\pm$ 0.01& 0.213  & $5.1^{+0.6}_{-0.5} $ \\
1ES 0414+009  &7.8$\pm$ 0.3  &  1.73$\pm$ 0.08 &  -0.04 $\pm$ 0.03 &0.287 &  --- \\
PKS 0447-439& 40.6$\pm$ 1.7  &  2.39$\pm$ 0.02&  0.14$\pm$0.01 &  0.316 & $13\pm 1.0$ \\
PKS 0548-322 & 3.30$\pm$0.15 &  2.08$\pm$0.08 &   0.09$\pm$ 0.04 & 0.373 & $130^{+50}_{-80}$ \\
PKS 0625-354 & 8.23$\pm$ 0.96 &  2.25$\pm$ 0.07 &  0.10 $\pm$ 0.02& 0.349  & $18^{+4}_{-5}$ \\
1RXS J101015.9-311909 & 9.15$\pm$ 0.42 &  2.30$\pm$ 0.11 & 0.20 $\pm$ 0.05 & 0.281 & $50^{+9}_{-8} $ \\
1ES 1101-232 & 16.1$\pm$0.1  & 1.70 $\pm$ 0.04 &  0.034 $\pm$ 0.02&  0.316 & $< 15.4\cdot 10^{3}$  \\
1ES 1312-423 & 5.94 $\pm$ 0.54 & 2.44 $\pm$ 0.17 &  0.37 $\pm$ 0.11 & 0.310  & $78^{+16}_{-13} $ \\
PKS 1440-389 &  43.8$\pm$1.0  & 2.36 $\pm$ 0.03& 0.16  $\pm$ 0.01& 0.259  & $18.6\pm 1.1$  \\
PKS 1510-089 & 20.0$\pm$2.5  & 2.88 $\pm$ 0.05 &  0.11 $\pm$ 0.01&  0.213 &  --- \\
AP Lib & 10.2 $\pm$ 0.9  & 2.51 $\pm$ 0.04&  0.09 $\pm$ 0.01 &  0.337 & $0.36^{+0.17}_{-0.11}$  \\
1ES 2322-40.9 & 25.4$\pm$2.1  &  2.12$\pm$ 0.07 & 0.12  $\pm$0.03 &  0.235 & $74^{+33}_{-23}$  \\
PKS 2005-489 & 17.4$\pm$0.1  &  2.29$\pm$ 0.02 & 0.13  $\pm$0.01 &   0.386 & $32.0^{+0.1}_{-0.1}$  \\
PG 1553+113 & 63.7$\pm$0.5  &  2.51$\pm$ 0.02 & 0.24  $\pm$0.01 &   0.287 & $36.1^{+0.1}_{-0.1}$  \\
RGB J0152+017& 9.7$\pm$1.8  &  2.17$\pm$ 0.05 & 0.09  $\pm$0.02 &   0.287 & $38.7^{+0.2}_{-0.1}$  \\
H 2356-309& 15.3$\pm$0.1  &  1.88$\pm$ 0.02 & 0.08  $\pm$0.01 &   0.287 & $1.6^{+1.3}_{-0.7} \cdot 10^3$  \\
1ES 0347-121& 9.7$\pm$1.8  &  2.17$\pm$ 0.05 & 0.09  $\pm$0.02 &   0.287 & $38.7^{+0.2}_{-0.1}$  \\
\hline

\end{tabular}
\end{center}
}
\tablefoot{The first column gives the name of the source. Data were fitted with a log-parabola model, corrected for EBL absorption. The normalisation $\phi_{\rm 0}(E_{\rm 0})$ and spectral parameters ($\alpha$ and $\beta$) are given. The fifth column gives the energy $E_{\rm 0}$ at which the normalisation is evaluated. The last column presents the calculated peak energy $E_{\rm peak}$ or the corresponding upper limit at 99\% CL in a SED representation. If no value is given, the peak or the upper limit cannot be computed (see text). Errors are statistical only.}
\end{table*}

\subsection{Spectral upper limits for selected objects}

H.E.S.S. observed several other extragalactic non-blazar objects during the time period covered by the HEGS dataset, but without any detection.
Table~\ref{table:UL} presents the integral flux upper limits at the 95\% CL. For the objects already targeted by H.E.S.S. and subject of a dedicated publication, no new observations are reported here and the updated limits are in agreement with those previously obtained. The main change in this work is the lower energy threshold reached thanks to the {\tt loose} cuts setting used in the analysis. We note that the limits reported in this work were obtained under the assumption of a point-like source and a spectral index of 3. 

Arp~220, at a distance of 77~Mpc \citep{2012ApJS..203...21A}, is an ultra-luminous-infrared galaxy, is characterised by high star-formation rates, making this object an interesting candidate for VHE $\gamma$-rays\ emission. In the HE range, the source 4FGL~J1534.7+2331 has been associated with Arp~220 in the 4FGL catalogue. The HEGS flux upper limit lies below the MAGIC ones \citep{Albert_2007} which makes these results more constraining for the hadronic emission models, further testing the emission mechanism \citep{2000A&A...362..937A,2005ApJ...620..244K}.

NGC~1068, at a distance of 14~Mpc \citep{2014MNRAS.440..696A}, is a local composite starburst and Seyfert 2 galaxy detected by \textit{Fermi}-LAT  and associated with 4FGL~J0242.6-0000. The origin of this giga-electronvolt gamma-ray emission is still under debate, with the LAT flux detected being due to either starburst or active galactic nucleus (AGN) activity. Comparisons of NGC 1068 with other starbursts have suggested that the $\gamma$-ray\ luminosity may be too high to be explained by the starburst activity alone \citep{2010A&A...524A..72L}. Future observations with the Cherenkov Telescope Array Observatory
\citep[CTAO,][]{2019scta.book.....C} are expected to offer sufficient sensitivity and spectral coverage so as to allow the starburst and AGN emission scenarios to be distinguished \citep{2019APh...112...16L}. The flux upper limits presented here, comparable to the limits obtained by the MAGIC collaboration \citep{2019ApJ...883..135A}, do not presently enable firm conclusions to be drawn. Interestingly, the recent hint of this source in multi-tera-electronvolt neutrinos at the 4.2$\sigma$ level by IceCube \citep{IceCube:2022der}, indicates an energy flux more than a factor of 10 brighter than the present tera-electronvolt gamma-ray upper limits. 

NGC~4945, at a distance of 3.8~Mpc \citep{2014MNRAS.440..696A}, is also a local composite starburst galaxy and Seyfert 2 system but the HE flux is consistent with starburst galaxy activity, and is not dominated by the active nucleus according to  \citet{2010A&A...524A..72L}. The recent data from the Pierre Auger Observatory seem to indicate that NGC~4945 could explain a hot spot of cosmic ray above 39~EeV \citep{Aab_2018}. With only 8.8 hours of data, the HEGS flux upper limit is incompatible with the an extrapolation of the flux detection by \textit{Fermi}-LAT.

Galaxy clusters were also observed to check for VHE emission.  Such emission could potentially be produced by Dark Matter annihilation from the cluster centre, such as Fornax~A, or by an AGN outburst, such as Hydra~A. In the case of Abell 496 and Abell 85, the data were used to probe the cosmic-rays production in the clusters. For theses objects, the analysis presented here reached a lower energy threshold than previously reported. For Abell 3376 or Abell 3667, no upper-limits were reported before with the H.E.S.S. telescope.

\begin{table*}
\small{
\caption{Integral flux upper limits on various observed objects.}
\label{table:UL}
\begin{center}
\begin{tabular}{c c c c c c}
\hline\hline 
Name  &  Type & Obs. Time & E$_{\mathrm{th}}$ &  Upper Limit  & References \\
& &  h & TeV & $10^{-12}\,{\rm cm^{-2}}\,{\rm s^{-1}}$ & \\\hline
NGC 4945 & Starburst/Seyfert 2   & 8.7 & 0.24 & 1.38 &  \\
NGC 1068  & Starburst/Seyfert 2 & 48.9 & 0.21 &  1.18 & 1 \\
Hydra A & Galaxy cluster &21.2 & 0.19 & 1.97 & 2 \\
Fornax A & Galaxy cluster  &  11.5 &0.20 & 0.80 & 3\\
Arp 220 & Starburst/Seyfert & 36.3 & 0.48 &  0.39 & 4 \\
Abell 3376  & Galaxy cluster &  7.4 & 0.22 & 1.78 &  \\
Abell 3667  & Galaxy cluster & 5.4 & 0.29 & 1.63 &  \\
Abell 496 & Galaxy cluster & 14.6 &0.23 & 0.93 & 5 \\
Abell 85  & Galaxy cluster & 39.0 & 0.20 & 1.00 &  6 \\
\hline

\end{tabular}
\end{center}
}
\tablefoot{The names and types of the objects are given, followed by the H.E.S.S. observation times in the third column, the achieved energy thresholds E$_{\mathrm{th}}$, and the Integral upper limits at a 95\% CL (above  E$_{\mathrm{th}}$). The last column provides references for previous results.
The Galaxy cluster Coma \citep{2009A&A...502..437A} is part of the HEGS dataset but due to its size, a point-like upper limit cannot be reported.}

\tablebib{
(1)~\cite{2019ApJ...883..135A}; (2) \citet{2012A&A...545A.103H}, (3) \citet{2012ApJ...750..123A}; (4) \citet{Albert_2007};
(5) \cite{2009A&A...495...27A}; (6) \cite{2009A&A...495...27A}.
}
\end{table*}

\subsection{Variability}
\label{variability}\label{sec:var}

\begin{figure*}
\centering
        \includegraphics[width=0.4\textwidth]{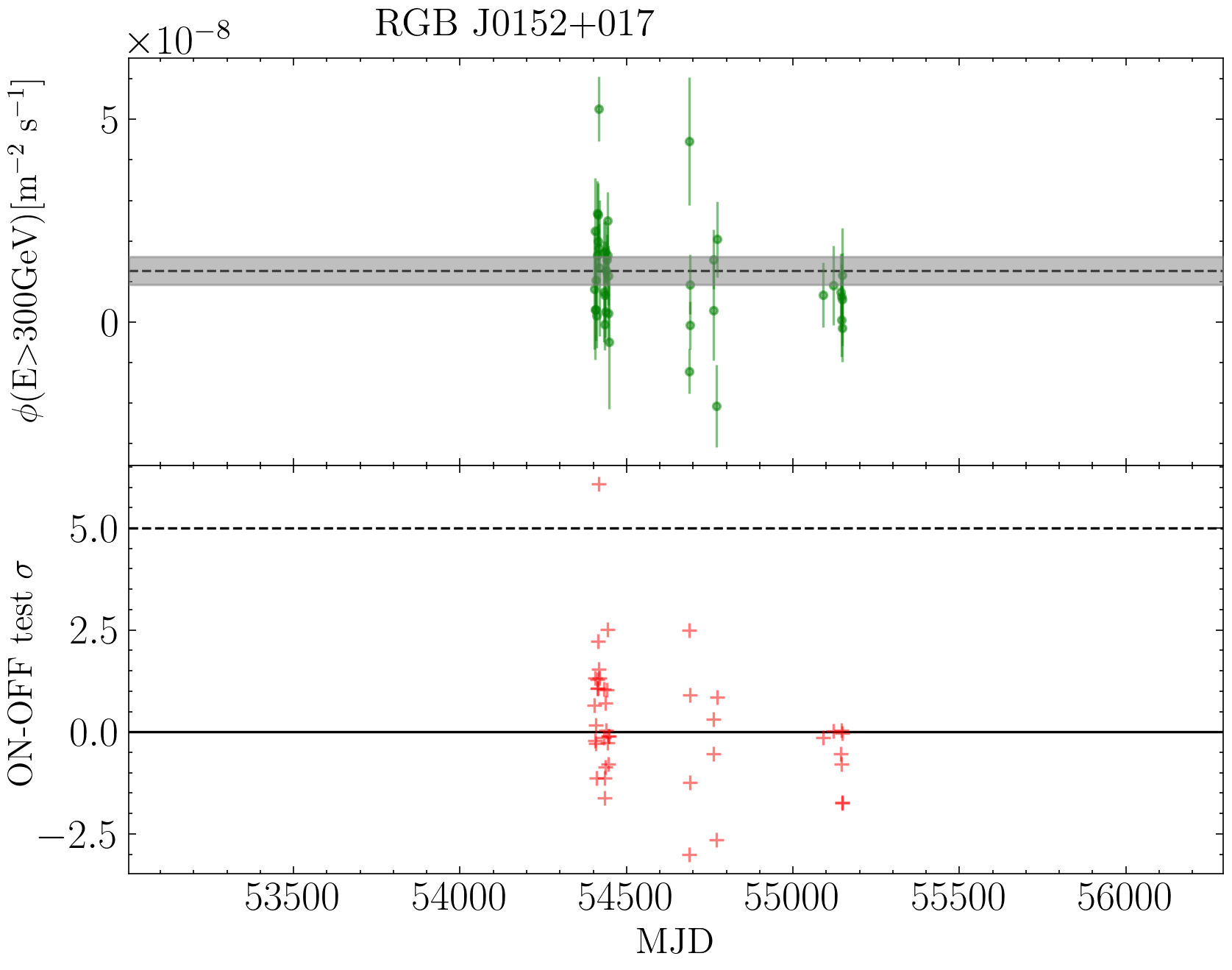}
        \includegraphics[width=0.4\textwidth]{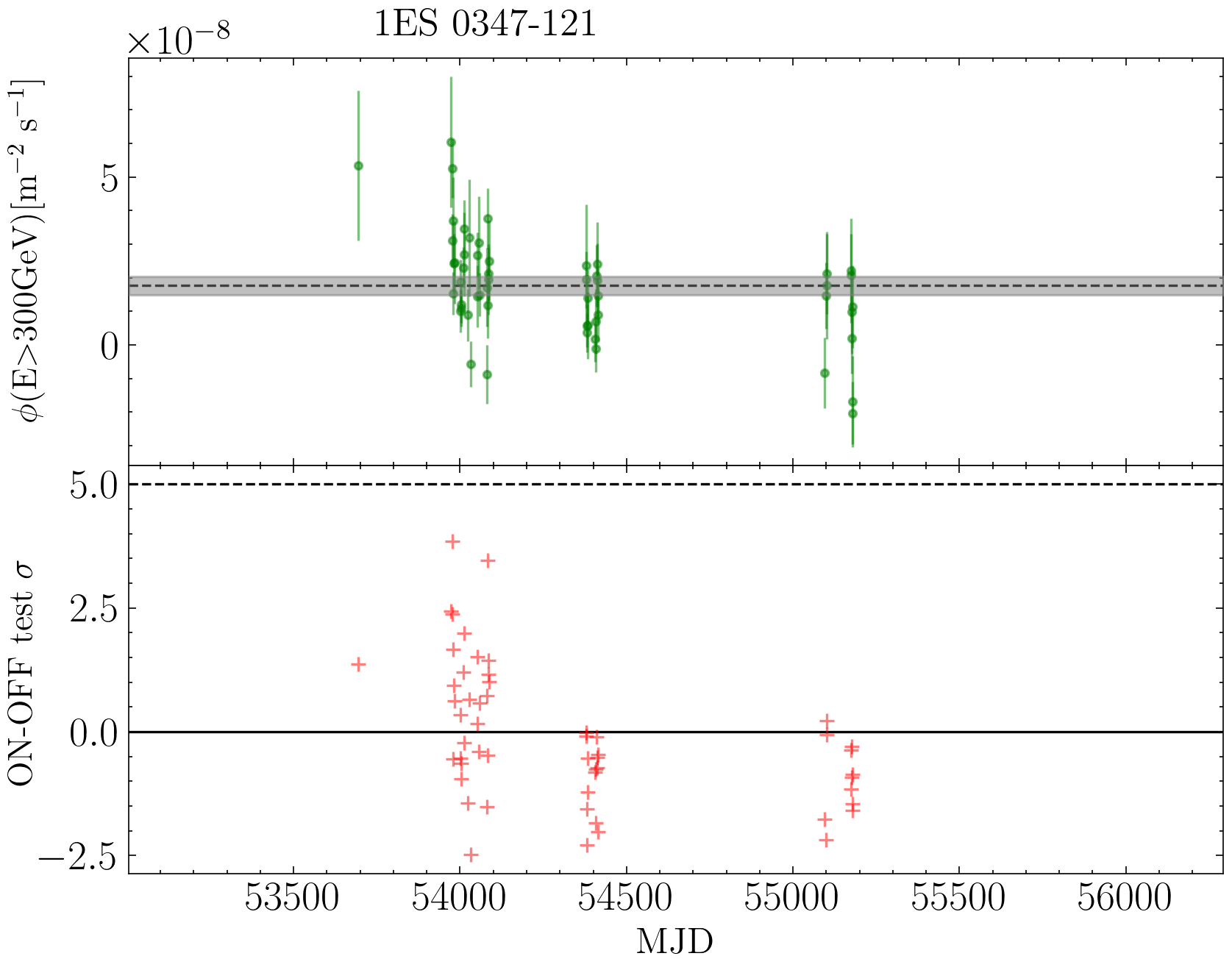}
        \includegraphics[width=0.4\textwidth]{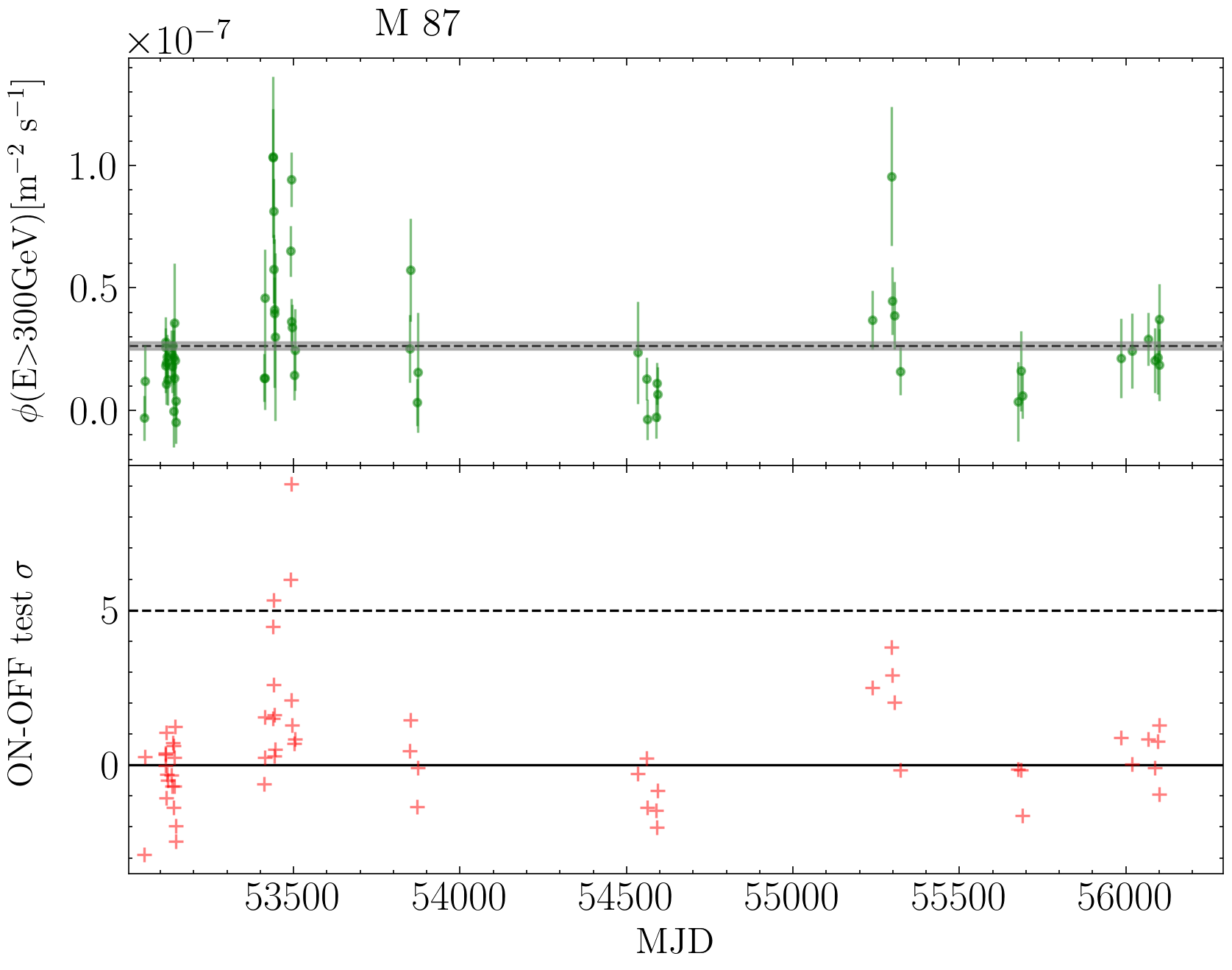}  
        \includegraphics[width=0.4\textwidth]{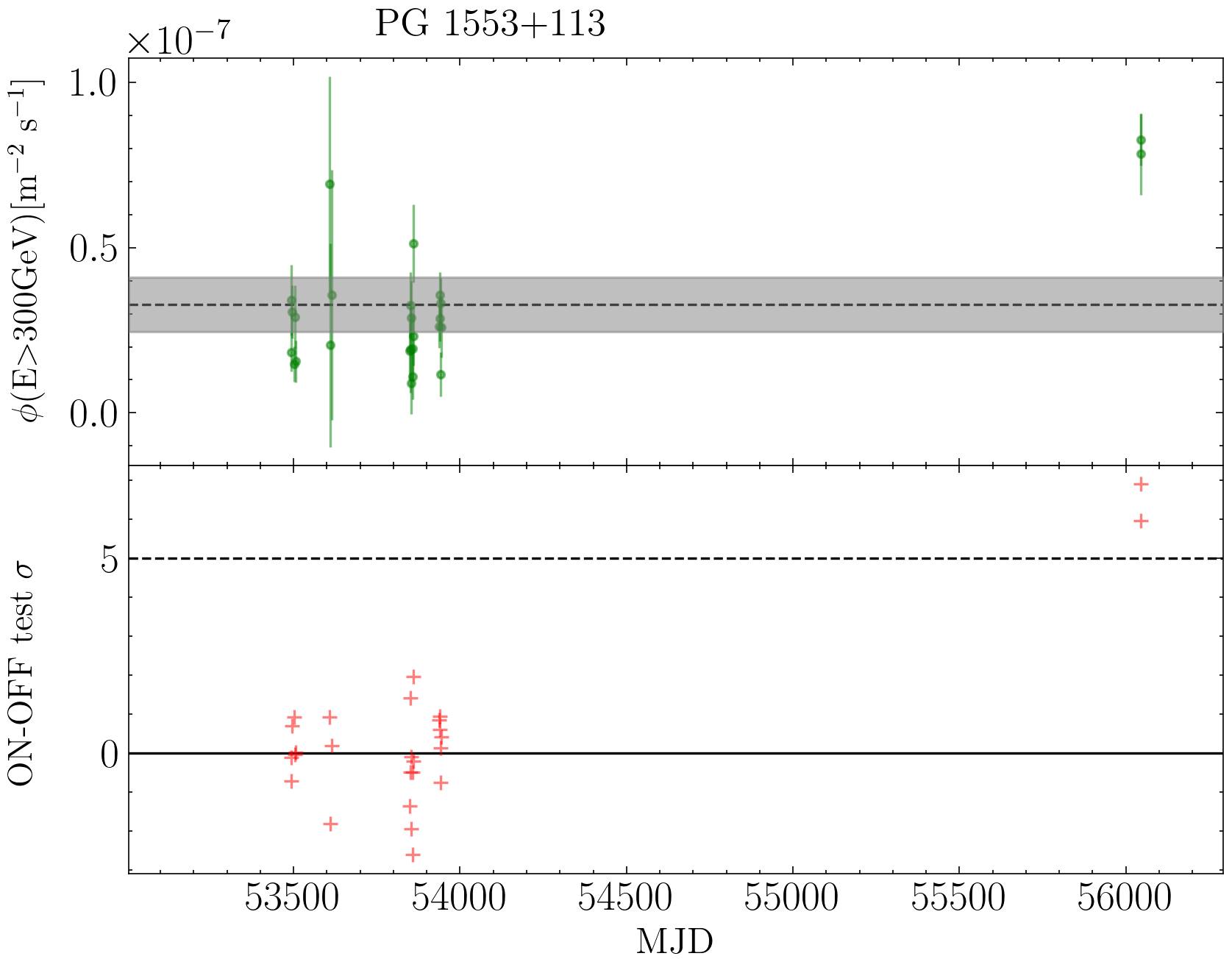}
        \includegraphics[width=0.4\textwidth]{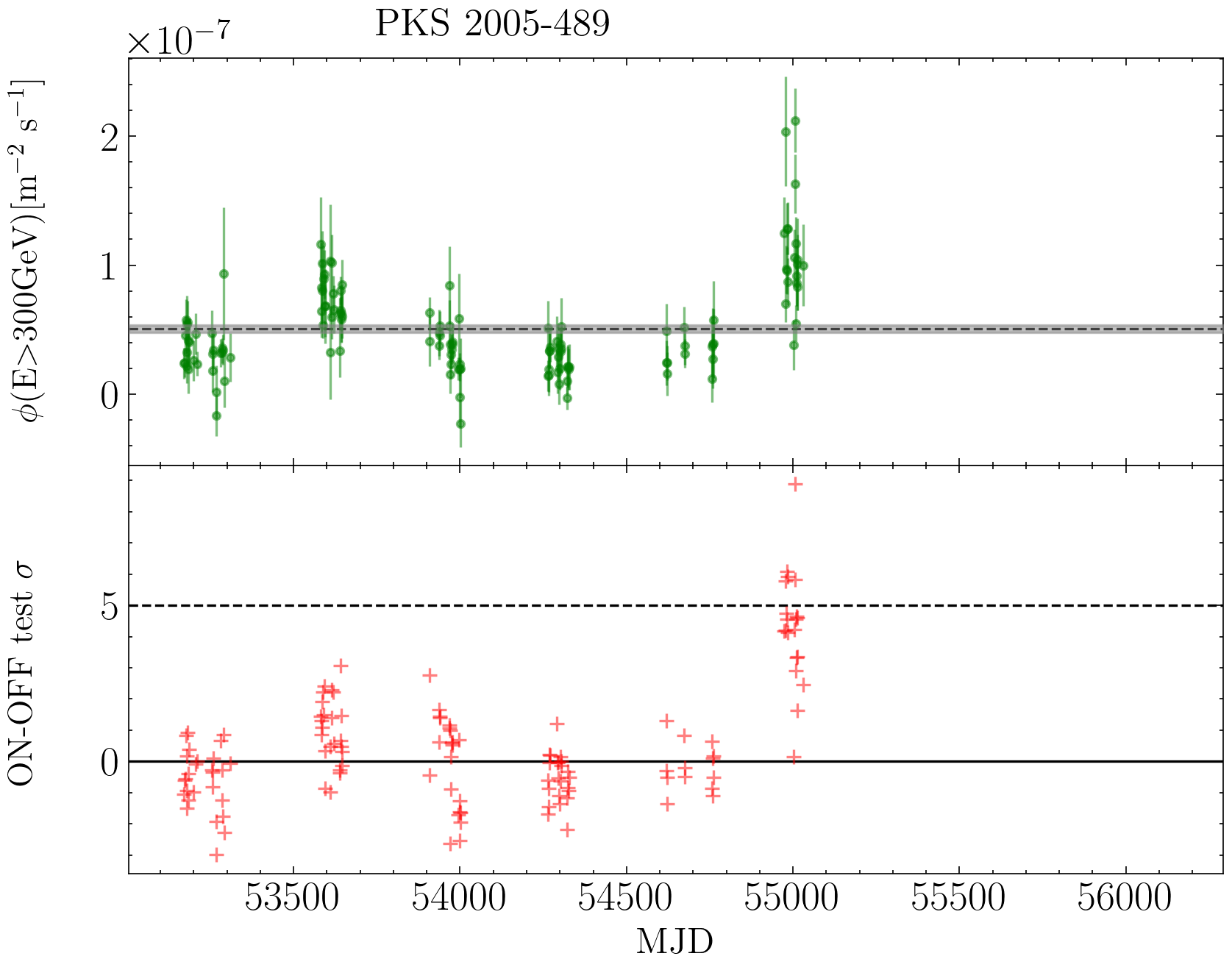}
        \includegraphics[width=0.4\textwidth]{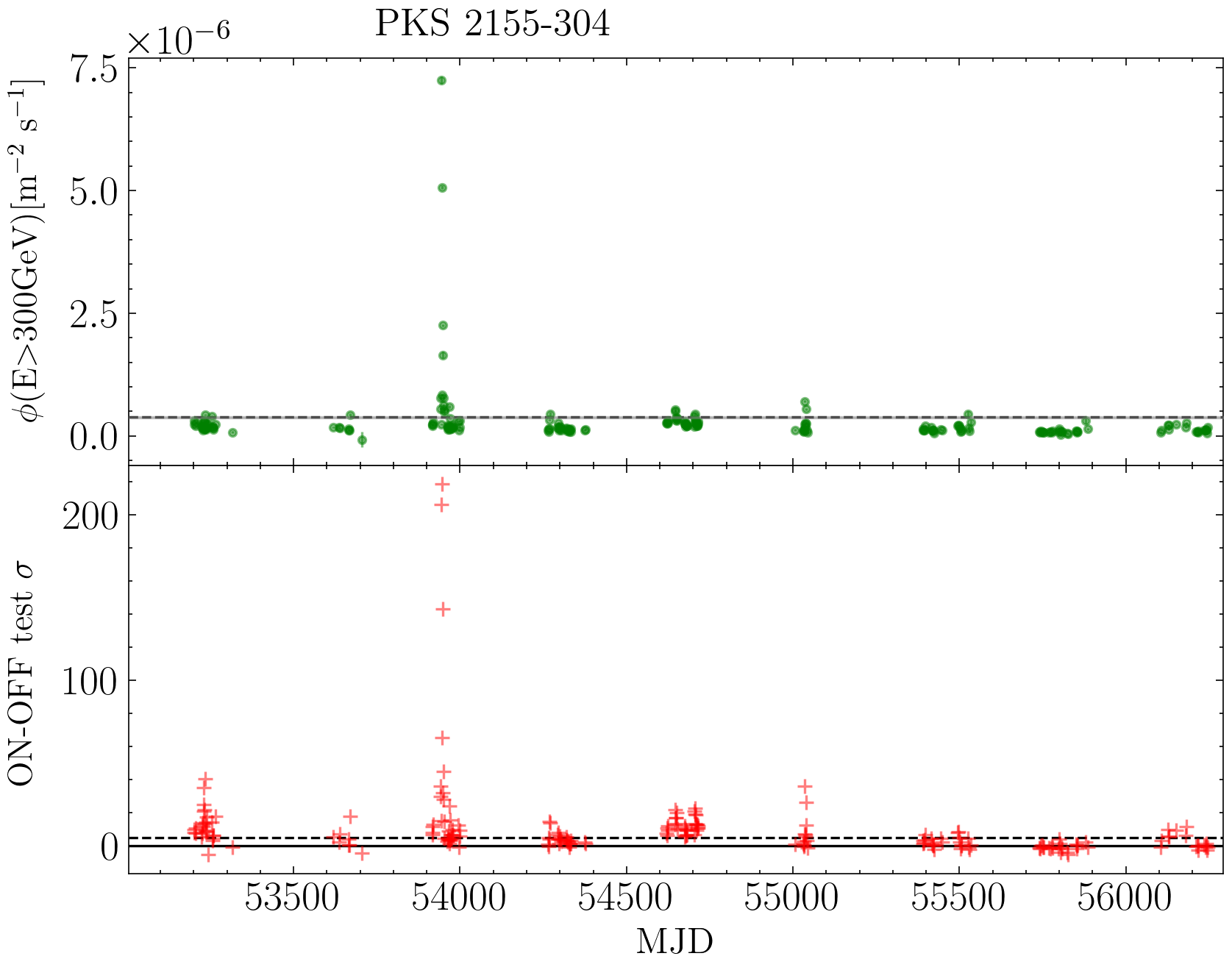}
        \includegraphics[width=0.4\textwidth]{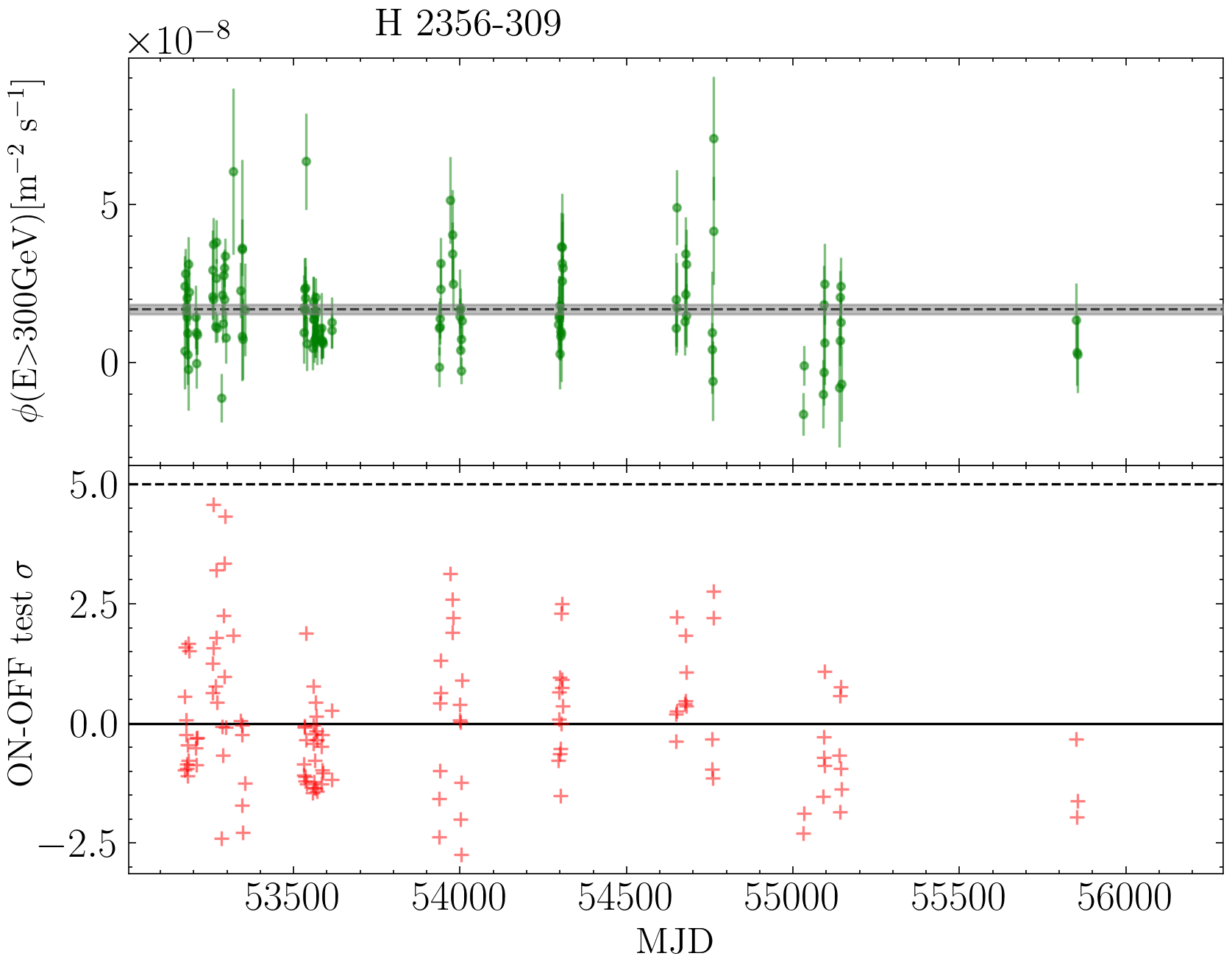}
    \caption{Light curves and  ON-OFF test significance. Top panel of each plot: Night-wise light curves of the sources found to be variable, derived using the best-fit spectral model as is described in Sect.~\ref{sub:spec}. The dashed line shows the mean flux level and the grey area shows the statistical uncertainties of this flux found from the spectral analysis. Lower panel of each plot: Temporal evolution of the ON-OFF Test significance for each source. The dashed line represents the 5$\sigma$ level.}
    \label{fig:lchegs}
\end{figure*}

Searches for variation in the flux across the observational time period were made in the manner described in Sect. \ref{meth:variability}: using the ON-OFF test, estimating the $\mathrm{F}_{\mathrm{var}}/\delta\mathrm{F}_{\mathrm{var}}$ ratio and by comparing the light curve to a constant. The results for the night-wise analysis are summarised in Table~\ref{table:variablesources} and the corresponding light curves are presented in Fig.~\ref{fig:lchegs}.

The bulk of sources with a positive ON-OFF test were already known to be variable sources and publications were made on their particular heightened activity states. This is the case for M~87 \citep{2012ApJ...746..151A}, PG~1553+113 \citep{2015ApJ...802...65A}, PKS~2005-489 \citep{2011A&A...533A.110H}. PKS~2155-304 is a very variable source which exhibited a prominent flare in 2006 \citep{2007ApJ...664L..71A}, showing variability on all timescales \citep[][]{2017A&A...598A..39H}, down to the shortest resolvable timescale of around a minute.

RGB~J0152+017 was not found to exhibit variability in the original publication. The increase in the observation time in the HEGS dataset (44.6 hours of live-time instead of 14.7 hours) allowed for the detection of variability with all tests.

For 1ES~0347-121, more data are presented in this study than in the original publication \citep{2007A&A...473L..25A}. The source is found to demonstrate variability, with a fractional variability value of $\mathrm{F}_{\mathrm{var}}  = 0.65\pm0.10$ (and $\mathrm{F}_{\mathrm{var}}/\delta\mathrm{F}_{\mathrm{var}}>5$). This object is possibly belonging to the extreme-HBL class \citep{2017arXiv171106282C} and the VHE flux variability might change the conclusion on the extragalactic magnetic field strength \citep{2022MNRAS.516.5379P}.

H~2356-309 was found to be variable on timescales of one month \citep{2010A&A...516A..56H}. In this study, we observe that it is also variable on a shorter timescale, that is on a nightly timescale, as is indicated in Table~\ref{table:variablesources}.

The Flat Spectrum Radio Quasar PKS~1510-089 was observed in a flaring state by MAGIC in 2015 \citep{2017A&A...603A..29A} and  H.E.S.S. and MAGIC in 2016 \citep{2021A&A...648A..23H}. For the H.E.S.S.-I time period considered in this study, the source did not show any evidence of variability.

For the high-frequency peaked BL Lac object (HBL) 1ES~0229+200, a preliminary analysis of the H.E.S.S. data previously indicated a hint of source variability \citep{2015ICRC...34..762C}. Here with almost the same dataset, no confirmation of this earlier hint of source variability is found.

\begin{table}
\small{
\caption{List of sources found to be variable in HEGS.}
\label{table:variablesources}
\begin{center}
\begin{tabular}{c c c c}
\hline\hline 
Name &On-Off test &$\chi^2$ Significance & $\mathrm{F}_{\mathrm{var}}$ \\\hline
RGB J0152+017 & yes & 4.89  & 0.82 $\pm$ 0.17 \\
1ES 0347-121  & no  & 5.13  & 0.65 $\pm$ 0.10  \\
M 87          & yes & 8.34  & 0.74 $\pm$ 0.08\\
PG 1553+113   & yes & 6.84  & 0.43 $\pm$ 0.11  \\
PKS 2005-489  & yes & 13.08 & 0.63 $\pm$ 0.04  \\
PKS 2155-304  & yes & -     & 2.24 $\pm$ 0.01 \\
H 2356-309    & no  & 6.88  & 0.51 $\pm$ 0.09  \\

\hline 
\end{tabular}
\end{center}
}
\tablefoot{The first column indicates the source names, the second indicates if the ON-OFF test gave a positive result. The last columns give the significance of the source variability from a fit with a constant on a night-by-night scale and the value of $\mathrm{F}_{\mathrm{var}}$. }
\end{table}

\subsection{log N-log S distribution}

\begin{figure*}
\centering
        \includegraphics[width=0.48\textwidth]{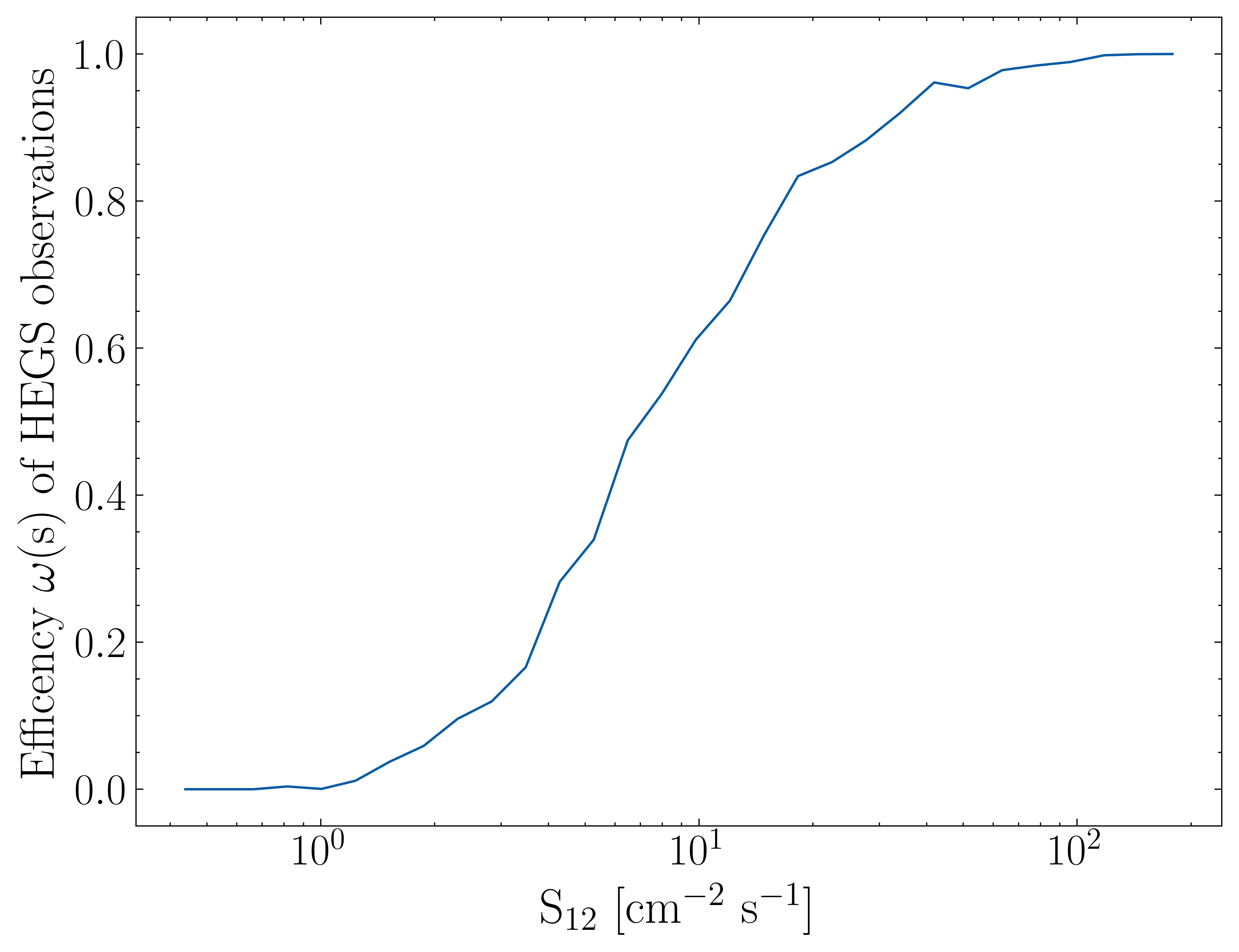}
        \includegraphics[width=0.51\textwidth]{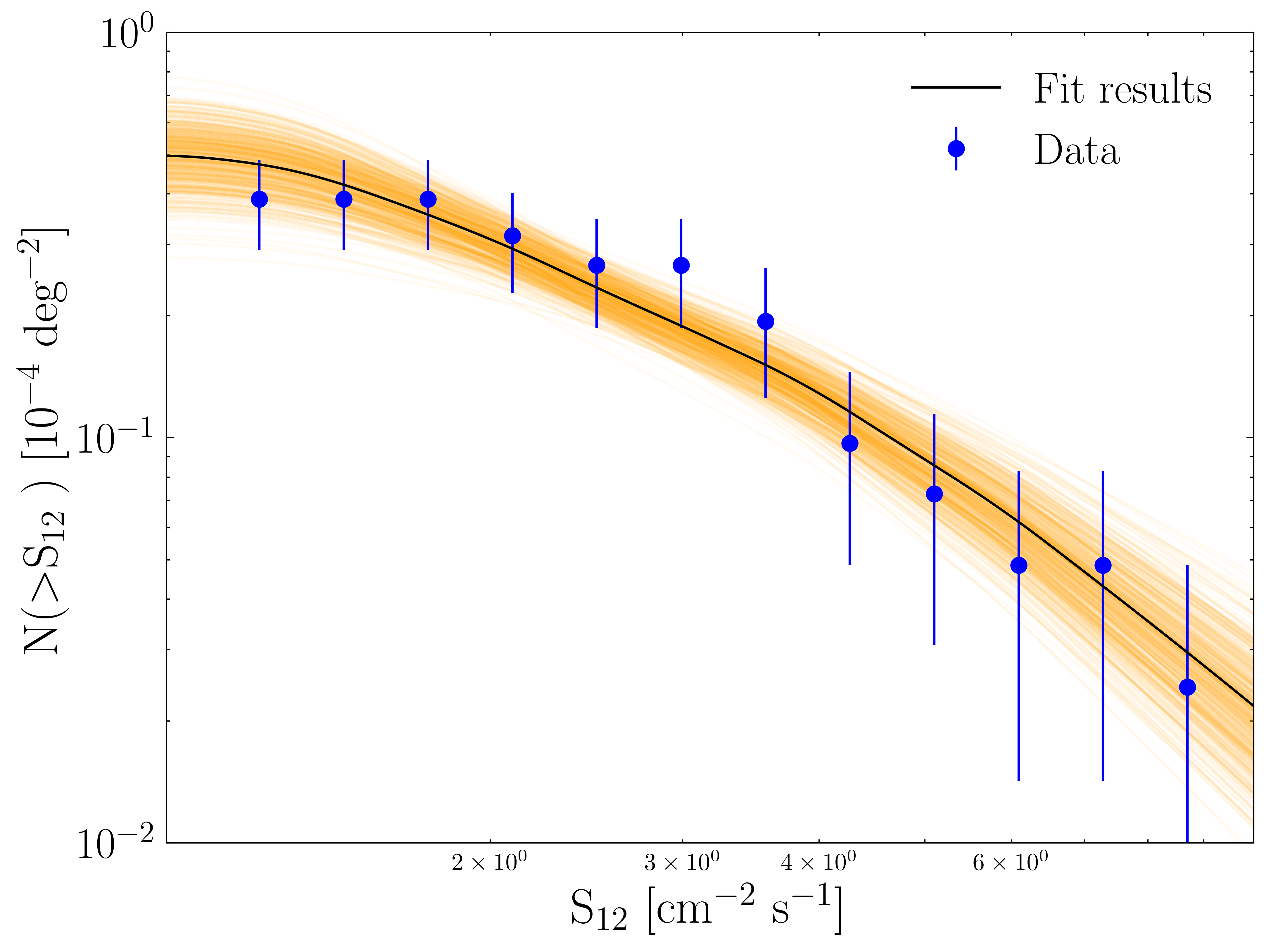}
    \caption{Source detection efficiency and log N-log S distribution. Left: Source detection efficiency, $\omega(S)$, as a function of the source flux $S_{12}$, in units of $10^{-12} cm^{-2} s^{-1}$, in the HEGS FoV. Right: Number of sources, $N$, above the flux $S_{12}$ as a function of $S_{12}$. The black line shows the best-fit power law model with index $\gamma=2.58^{+0.23}_{-0.24}$ convolved with the H.E.S.S. efficiency and the orange area shows the error.}
    \label{fig:lognlogs}
\end{figure*}

The efficiency for the detection of sources with the HEGS observational dataset has been estimated using a sampling simulation method. Sources with a random level of flux, and placed at random positions with non-zero exposure, were simulated. For each sky position, the observed number of OFF events, the observation time, and the energy threshold were computed (see Sect.~\ref{sec:maps}). The simulated flux was subsequently converted to a number of simulated ON events and the corresponding significance \citep{Lima} was computed. Comparing the results from this to the sensitivity at the simulated position, one can decide whether such a simulated source would be detectable or not. The results from this study, namely the detection fraction as a function of the flux, also called the detection efficiency $\omega(S)$, is shown in the left-panel of Fig.~\ref{fig:lognlogs}.

The cumulative number of BL~Lac\ objects, $N$, above the gamma-ray flux, in units of $10^{-12} \mathrm{cm}^{-2} \mathrm{s}^{-1}$, $S_{12}$  is shown in the right panel of Fig.~\ref{fig:lognlogs}. In the sample presented here, the sources found to be variable were removed so as to not bias the results towards a high flux value.

Ignoring any evolution effect, the number of sources per energy flux bin, $dS$, can be approximated by a power law, $\frac{dN}{dS}\propto S^{-\gamma}$ \citep[see][]{2010ApJ...720..435A}. The fit of the $N(>S)-S$ distribution with such a power law has been performed taking into account the efficiency $\omega(S)$.  The observed number of sources $N$ above flux $S$ is then:

\begin{equation}
    N_{observed}(>S) = \int_{S}^{\infty}\omega(S') \frac{dN}{dS'}dS'.
\end{equation}

The fit to the brightness distribution data was performed using a Markov chain Monte Carlo (MCMC) algorithm. The result from the MCMC scan is presented in the right panel of Fig.~\ref{fig:lognlogs}.  The best-fit index is found to be $\gamma=2.58^{+0.23}_{-0.24}$. This
value for the index is in agreement with the Euclidean expectation of $2.5$, but a value below this level is naturally obtained once the effects of both the attenuation of flux by the EBL and source evolution effects of a subset the BL~Lac class (see Sect.~\ref{class_bias}) are
taken into account. In both cases, a suppression of the flux from more distant sources is brought about. To appreciate this effect,
the disappearance of more distant sources can be described by the parameter $\alpha$, with the number
of sources with distance $R$ scaling as $N\sim R^{3-\alpha}$. Since $S\propto R^{-2}$, the introduction of the parameter $\alpha$ leads to a source brightness distribution following $dN/dS\propto S^{-(5-\alpha)/2}$. A value of $\alpha=0.2$ has been found by \citet{2014ApJ...780...73A}, which yields a value of $\gamma=2.4$. Both the value of 2.4, obtained with the diminution of flux with redshift, and the value of 2.5 obtained without such effects, are in agreement with the fit values we obtain. We note that this estimate ignores any source brightness variations due to jet Doppler effects, as well as the variability of the sources.

\section{Discussion of the HEGS results}\label{HEGSInterp}
\subsection{Source class discussion}

\label{Sources}

A variety of extragalactic source objects can be detected by H.E.S.S.; namely, starburst galaxies, RGs, and blazars. Due to the particularity of the H.E.S.S. data taking, GRBs are not discussed in this work and a forthcoming paper using the H.E.S.S.-I data will be presented. Blazars, being the only beamed emitter detected in this analysis, are sufficiently bright so that their emission can be detected from relatively large distances in the VHE $\gamma$-ray\ band \citep[up to a redshift of at least $z=0.997$,][]{2023ATel16381....1C}. The population of blazars itself is commonly further decomposed into FSRQ and BL~Lac\ objects, with the former possessing larger intrinsic luminosities than the latter which emit up to higher energy. A complete breakdown of the objects detected by H.E.S.S. is provided in Table~\ref{table:MeanIndex}.

The FSRQs and BL~Lac\ objects can also be classified using the low energy peak position. Blazars can then be classified low-synchrotron-peaked (LSP) blazars, intermediate-synchrotron-peaked (ISP) blazars, or as high-synchrotron-peaked (HSP) blazars \citep{Ajello_2020}. Most of the  BL~Lac\ objects in HEGS are HSP objects, except for AP~Librae.

The large majority (18 out of 23) of the sources in the HEGS catalogue belong to the BL~Lac\ object class. This class of objects presents a high flux in X-rays, an indication of the presence of very high energy electrons in the source, which in turn can also produce VHE photons. Consequently, such objects have been a primary extragalactic target for Cherenkov telescope observations. This observational focus at least partly explains why the BL~Lac\ class is so abundant in the set of objects detected within the HEGS catalogue. The mean EBL-corrected VHE spectral index for the BL~Lac source class in the HEGS catalogue is 2.16 (see Table~\ref{table:MeanIndex}) with a standard deviation of the distribution of 0.47. 

The only FSRQ present in the HEGS catalogue is PKS~1510-089. The analysis covers the HE flare from this object in March 2009 reported in \citet{2013A&A...554A.107H} and data taken in July 2010, July 2011 and May 2012. The source is known to be variable in the VHE band \citep{2017A&A...603A..29A,2021A&A...648A..23H}.  The reported flux in the HEGS catalogue is of the same order as the flux reported by MAGIC above 175~GeV  \citep{2018A&A...619A.159M} or H.E.S.S. \citep{2023ApJ...952L..38A} but with a spectra softer than the one reported in the latter publication.

Two of the closest RGs are detected in the HEGS catalogue: Centaurus~A and M~87, situated at distance of 3.8~Mpc \citep{1992AJ....103...11F} and 16.8~Mpc \citep{2019ApJ...875L...6E}, respectively. Detailed discussion of each object can be found in \citet{2018Galax...6..116R}. In the presented dataset, Centaurus~A is detected at 12.0$\sigma$, exhibiting steady weak emission in the VHE band. M~87 is found to be variable during the H.E.S.S. observation at a 6.8$\sigma$ level on a nightly timescale. The classification of the object PKS~0625-35 remains a subject of continued debate, with some classifying it as a RG, and others as a blazar \citep{2018Galax...6..116R}. The object presented both BL~Lac and RG-like behaviour in the multi-wavelength data gathered during a VHE flare \citep{2024A&A...683A..70H}. The average EBL-corrected VHE spectral index for the RG source class in the HEGS catalogue is 2.43 (see Table~\ref{table:MeanIndex}).

The only starburst galaxy present is NGC~253, situated at a distance of 3.5~Mpc \citep{2005ApJS..160..149S}. The source is detected at 7.0$\sigma$. The reported analysis reveals a hard spectral index, with $\Gamma=2.34$. No evidence for variability in the emission from the source is found.

\subsection{The fourth \textit{Fermi}-LAT catalogue in the HEGS FoV}

\subsubsection{Spectral constraints on fourth \textit{Fermi}-LAT  catalogue  sources}
\label{sec:Fermicat}

In the HE range, the fourth \textit{Fermi}-LAT  catalogue of sources \citep[4FGL;][]{2019arXiv190210045T} can be used to search for potential VHE emitters. In total, 247 sources are in the HEGS FoV. For each point in the sky observed by H.E.S.S., an energy threshold, $E_{\rm th}$, can be defined that is dependent on the observational conditions (zenith angle of the observation, telescopes participating, etc.) . To compare the spectrum measured by \textit{Fermi}-LAT  and the HEGS observations, the best-fit spectrum reported in the 4FGL was extrapolated above $E_{\rm th}$ and corrected for EBL attenuation using the \citet{2011MNRAS.410.2556D} EBL model. The extrapolated integral flux is computed for each 4FGL object in the HEGS FoV. A \textit{Fermi}-LAT  source is classified as constrained if this flux is higher than the HEGS  upper-limit on the integral flux, obtained assuming a default spectral index of 3.

To correct for the EBL absorption, the redshift quoted either in the 4th \textit{Fermi}-LAT  catalogue of Active Galactic Nuclei \citep[4LAC, ][]{Ajello_2020} or in the Roma BZCAT \citep{2015Ap&SS.357...75M} or in the Third \textit{Fermi}-LAT  Catalogue of High-Energy Sources \citep[3FHL, ][]{2017ApJS..232...18A} was used. Following the same procedure adopted in the 4LAC, a default redshift of $z=0.3$ was assumed when the source redshift was unknown.

\begin{figure*}
\centering
        \includegraphics[width=0.95\textwidth]{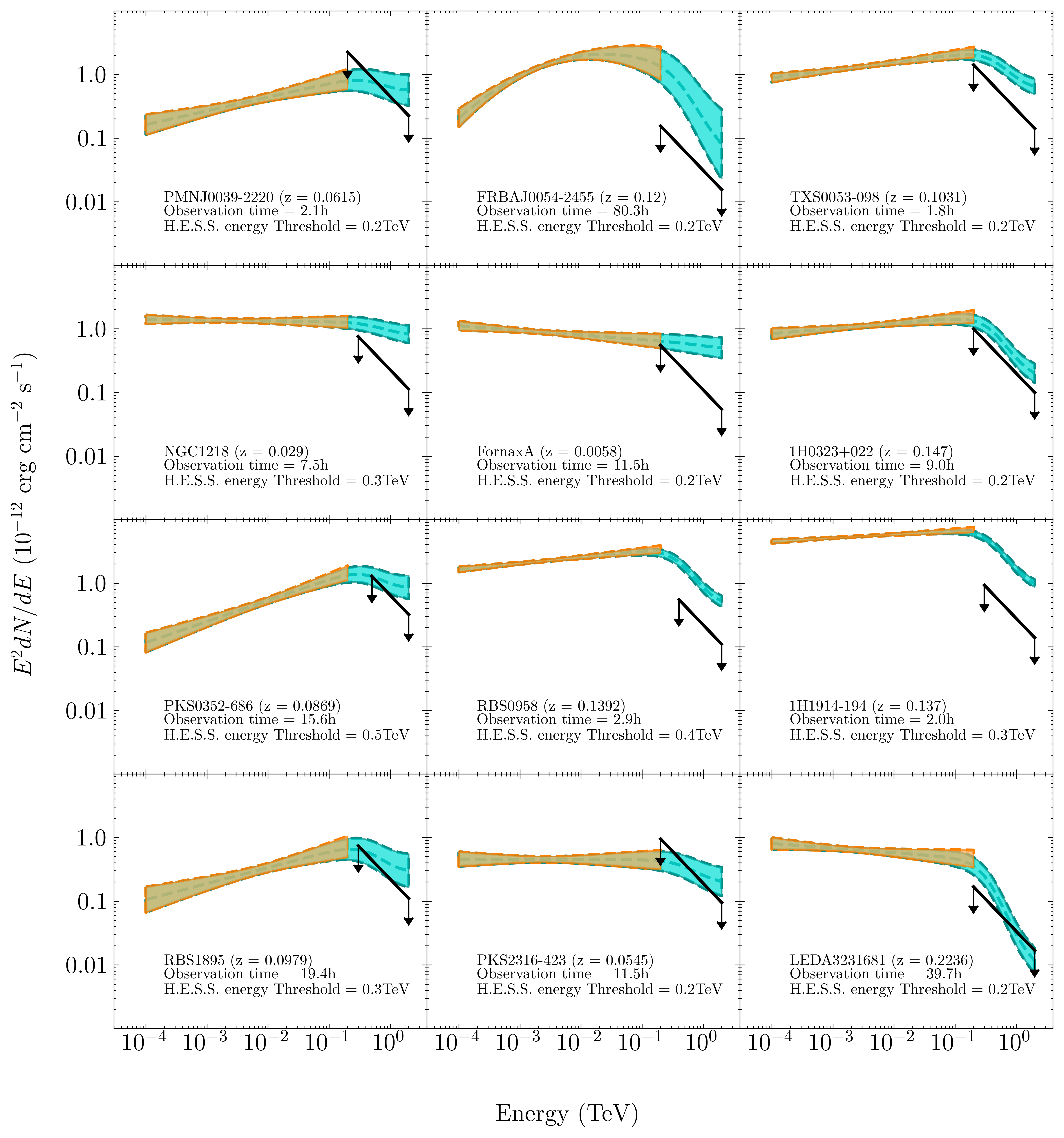}
    \caption{Spectral energy distributions of the 12 sources constrained by the observations presented in this work. The orange butterfly represents the \textit{Fermi}-LAT measurement (from the 4FGL), and the cyan one indicates the extrapolation towards the energies observable by H.E.S.S. including EBL absorption. The black line represents the observed H.E.S.S. upper limit at 95\% CL assuming an index of 3.}
    \label{fig:fermiHEGSseds}
\end{figure*}

Of the 247 \textit{Fermi}-LAT  VHE source candidates, an extrapolation of their spectra found that 12 of them are constrained by the H.E.S.S. observations.
The list is given in  Table~\ref{table:constrainedsources}. As is shown in Fig.~\ref{fig:fermiHEGSseds}, an extrapolation of the HE emission into the VHE domain predicts a flux higher than the HEGS upper limit. Four objects in the list were reported as variable in the 4FGL with a variability index greater than 18.48 (NGC 1218, RBS 0958, 1H 1914-194 and LEDA 3231681). A visual inspection of the \textit{Fermi}-LAT public light-curves\footnote{\url{https://fermi.gsfc.nasa.gov/ssc/data/access/lat/LightCurveRepository/index.html}} do not reveal structure that can bias the flux.

RBS0958 and 1H1914-194 are the brightest objects in this sample. While observed only for several hours, the extrapolation of the \textit{Fermi}-LAT  spectra is almost a factor 4 higher than the HEGS measurement. For all constrained objects, the existence of a break feature in the source spectrum is required in order to be compatible with their non-detection in the VHE band.

\begin{table*}
\small{
\caption{List of sources constrained by the HEGS observations.}
\label{table:constrainedsources} 
\begin{center}
\begin{tabular}{ccccccccc}
\hline\hline 
4FGL Name & Other Name & Type &  ObsTime &$E_{\rm th}$ & $\sigma$& $\phi_{\rm 1 TeV}$\\
&& &  hours &  TeV &    & $10^{-13}$ cm$^{-2}$\,s$^{-1}$\,TeV$^{-1}$\\    \hline
4FGL J0039.1-2219  & PMN J0039-2220 & BL Lac & 2.1 & 0.2 & 1.3  & 2.81\\ 
4FGL J0054.7-2455 & FRBA J0054-2455 & BL Lac & 80.3 & 0.2 & -1 & 0.19\\ 
4FGL J0056.3-0935  & TXS 0053-098 & BL Lac & 1.8 & 0.2 & -0.4  & 1.77\\ 
4FGL J0308.4+0407 & NGC 1218 & RG & 7.5 & 0.3 & 0.2 & 1.41 \\ 
4FGL J0322.6-3712e & Fornax A & RG & 11.5 & 0.2 &0.2  & 0.68\\ 
4FGL J0326.2+0225 & 1H 0323+022 & BL Lac & 9.0 & 0.2 & 0.8 &  1.25\\ 
4FGL J0353.0-6831 & PKS 0352-686 & BL Lac & 15.6 & 0.5 & 2.0 &  4.03\\ 
4FGL J1117.0+2013 & RBS 0958 & BL Lac & 2.9 & 0.4 & -2.1 &  1.38\\ 
4FGL J1917.7-1921 & 1H 1914-194 & BL Lac & 2.0 & 0.3 & -0.4  & 1.75\\ 
4FGL J2246.7-5207 & RBS 1895 & BL Lac & 19.4 & 0.3 & 1.2  &  1.39\\ 
4FGL J2319.1-4207 & PKS 2316-423 & BL Lac & 11.5 & 0.2 & 0.9  & 1.19\\ 
4FGL J2350.6-3005 & LEDA 3231681 & BL Lac & 39.7 & 0.2 &-1.4  & 0.21\\ 

\hline 
\end{tabular}
\end{center}
}
\tablefoot{The table gives the actual 4FGL name of the source and its association. The next columns give the sources type, the observation time, the corresponding energy threshold and the significance. The last column is the upper limit at 1 TeV (95\% CL). The positions used to extract the values are those from the 4FGL catalogue.}
\end{table*}

\subsubsection{Connecting the \textit{Fermi} and H.E.S.S. skies}\label{simu}

The \textit{Fermi}-LAT and H.E.S.S. energy ranges are very close. It is therefore natural to compare the results of both instruments and explore if the same luminosity function, derived in the \textit{Fermi}-LAT energy range, also describes well the HEGS observations. To achieve such a comparison, numerical simulations of the $\gamma$-ray~sky are performed. The blazar population and its cosmological evolution is described by the luminosity function $\Phi(L_\gamma,z,\Gamma)$ which gives the number of sources as a function of the luminosity $L_\gamma$, the redshift $z$, and the spectral index $\Gamma$. Using \textit{Fermi}-LAT data, the luminosity functions for the FSRQs \citep{2012ApJ...751..108A}, BL~Lac\ \citep{2014ApJ...780...73A} separately and jointly together \citep{2015ApJ...800L..27A} had been evaluated in the \textit{Fermi}-LAT energy range (100~MeV-100~GeV).

The spectral index $\Gamma$, used in the simulation, was evaluated using \textit{Fermi}-LAT catalogues. The 4LAC based on the 4FGL reports a mean spectral index of $2.02\pm0.21$ for BL~Lac objects and $2.44\pm0.20$ for FSRQs, in the 100 MeV to 100 GeV energy range. The same values can be extracted from the 3FHL but above 10~GeV. In this energy range, the BL~Lac mean index is $2.46 \pm 0.52$ and for FSRQs, the mean index is $3.25 \pm 0.76$.  Thus, the \textit{Fermi}-LAT catalogues indicate the presence of a spectral break of 0.4 for the BL Lac objects and 0.8 for the FSRQs. These values are used in the rest of the paper, since the results provided by the \textit{Fermi}-LAT catalogues, based on a larger source sample, are both consistent and also more statistically significant, than the results obtained from the HEGS analysis (which obtained $\Delta \Gamma_{{\rm BL LAC}} =0.33$, see Sect.~\ref{comparCat}).

Using the two luminosity functions for BL~Lacs and FSRQs, simulations of the sky were performed with a MCMC assuming that the \textit{Fermi}-LAT luminosity function holds in the HEGS energy range. The three variables ($L_\gamma,z,\Gamma$) are randomly drawn and the MCMC algorithm selects the triplets to build a realistic population.

An ensemble of $75\times 10^6$ sources with parameters $L_\gamma,z,\Gamma$ were drawn from both the BL~Lac and FSRQ luminosity functions as well as uniform random positions on the HEGS FoV. The spectrum of each simulated source was determined assuming a power-law shape with parameters determined from the luminosity, $L_\gamma$, redshift, $z$, and index, $\Gamma$. The luminosity, $L_\gamma$, expressed in the \textit{Fermi}-LAT energy range, can be converted to a normalisation at a given energy. It is assumed that this power-law shape is valid up to 10~GeV. At higher energies than this, the EBL correction is applied and an extra spectral break of $\Delta \Gamma = 0.4$ for BL Lac and $0.8$ for FSRQs is assumed. This allows the spectrum to be extrapolated into the H.E.S.S. energy range.

\begin{figure*}
\centering
        \includegraphics[width=0.49\textwidth]{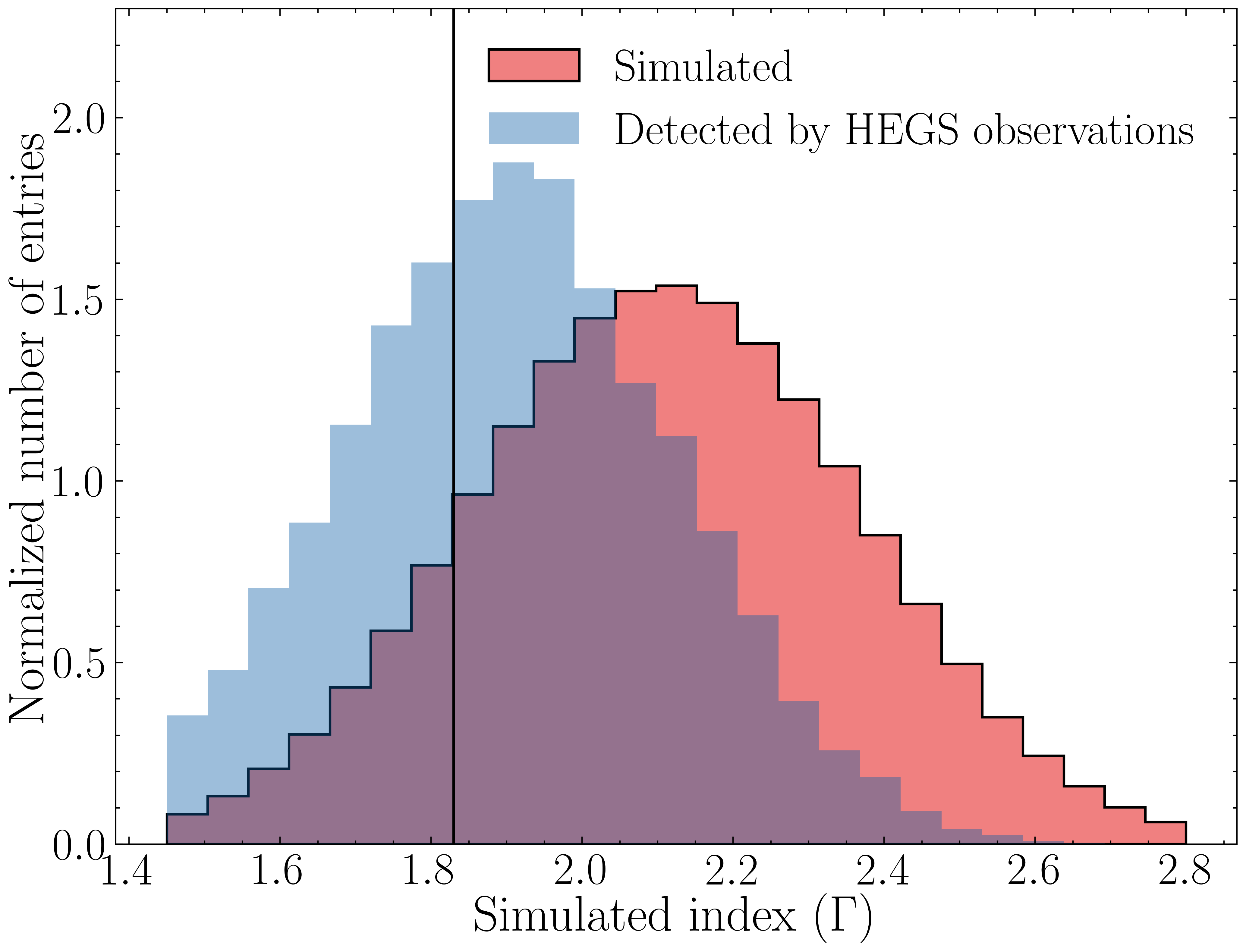}
        \includegraphics[width=0.49\textwidth]{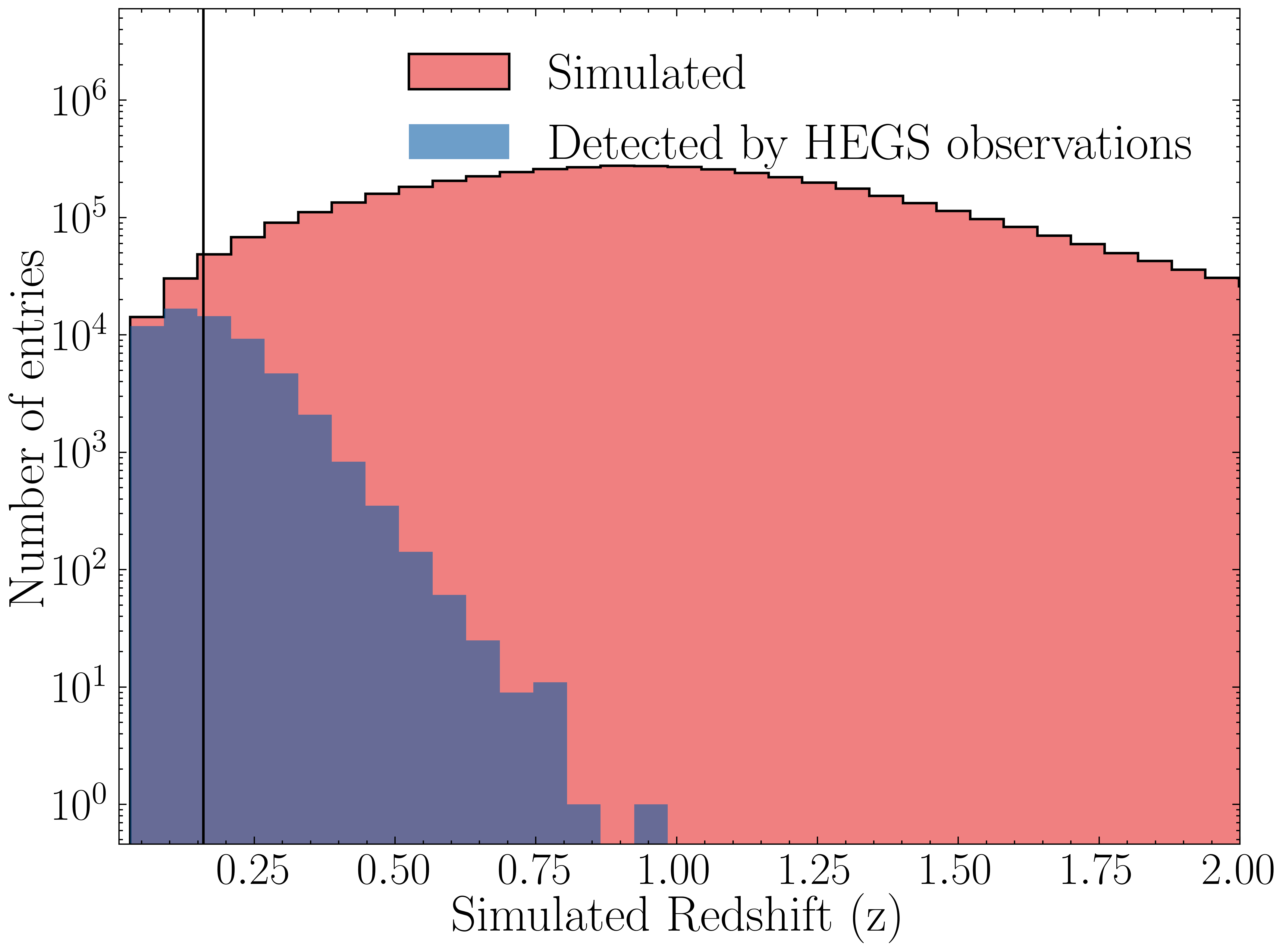}
    \caption{Simulated distribution of spectral index and redshift. Left: Distribution of the spectral indices of simulated BL Lac sources (red), along with simulated BL Lac that would have been detected with the HEGS observation (blue). For visibility purposes, both distribution are normalised to have a density of 1. Right:  Distribution of the redshift for all simulated BL Lac sources (red) and for the BL Lac expected to be detected (blue). In both graphics, the black line corresponds to the actual mean of the population of BL Lac objects.}
    \label{fig:simIndexRedshift}
\end{figure*}

Using the HEGS sensitivity map for a 5.7$\sigma$ detection, the expected distributions for both the detected source spectral index and redshift can be produced via simulations. A comparison of the observed distributions of these observables with those obtained via simulations are shown in Fig.~\ref{fig:simIndexRedshift}. For BL~Lac objects, the simulated mean index (in the \textit{Fermi}-LAT  energy range) of objects expected to be detectable by HEGS is 1.92 $\pm$ 0.05 and the mean redshift is 0.16 $\pm$ 0.01. This expected mean redshift value is close to the actual mean redshift reported for BL~Lacs in Table~\ref{table:MeanIndex}, and the expected mean spectral index is sightly harder than that observed. Overall, these simulations appear to account for the observational bias of the \textit{Fermi}-LAT  blazars at tera-electronvolt energies.

This comparison of the simulated and observed source parameters highlights the effect that the sensitivity of the instrument, along with the impact of the EBL, together play on the observational bias. Harder and closer sources are found to be preferentially detectable, an effect which must first be well understood in order that the compatibility of the data with the underlying blazar luminosity function can be evaluated.

Having simulations of the blazars with detection rates in hand, it is possible to compute the number of sources that should be detected serendipitously in the HEGS FoV, allowing one to compare and contrast this estimate with the actual rate of serendipitous detections in the VHE range. With this aim, the number of simulated sources is normalised to match the total number of 3FHL sources. In detail, a flux threshold of $4\times10^{-9}\ \mathrm{photons} \cdot$cm$^{-2}$\,s$^{-1}$ was used in our simulation, resulting in 750 BL~Lac and 172 FSRQ sources having a flux above the threshold in the 3FHL catalogue. Subsequently adopting this same normalisation factor for the simulation in the VHE range, 2.07 $\pm$ 0.09 BL~Lacs objects and 0.06  $\pm$  0.01 FSRQs would have been expected to be detected serendipitously within the HEGS observations; that is, without dedicated observation. This estimate is in agreement with the 2 BL~Lac\ discoveries made by H.E.S.S. which are 1ES~1312-423 \citep{2013MNRAS.434.1889H} and 1ES~2322-40.9 \citep{2019MNRAS.482.3011A}. Those two objects were in the FoV of other H.E.S.S. targets and their detections were achieved based on non-dedicated observations.

\subsubsection{Observational bias of blazar classes}
\label{class_bias}

An estimate of the HEGS observational bias to preferentially detect BL~Lac objects over FSRQs can be additionally evaluated with the aid of simulations. The ratio of FSRQs to BL~Lacs found in catalogues is strongly dependent on the threshold energy of the catalogue. This is demonstrated by consideration of the 3LAC \textit{Fermi} catalogue \citep{2015ApJ...810...14A}, which had a threshold energy of 100~MeV, and found a BL~Lac to FSRQ ratio of 0.3. Conversely, the \textit{Fermi} 3FHL catalogue had a threshold energy of 10~GeV, and found a different BL~Lac to FSRQ ratio of 4.4.

Applying our simulation up to energies of several tera-electronvolts, a clear preference for the detection of BL~Lac objects is found with a BL~Lac to FSRQ ratio of 40 being obtained. 
These results are consistent with the serendipitous blazar discoveries
found from HEGS, which were solely of the BL~Lac type.
It should be additionally noted that due to the threshold energy of the instrument, H.E.S.S. observations
of BL~Lac blazars preferentially detect the HBL population sub-class of the BL~Lac source class. This subclass of BL~Lac classifies sources for which the frequency of the peak of the synchrotron emission, $\nu_{\rm peak}$, in the SED sits above $10^{15}$~Hz \citep{2010ApJ...716...30A}.

\begin{figure}
\centering
        \includegraphics[width=0.45\textwidth]{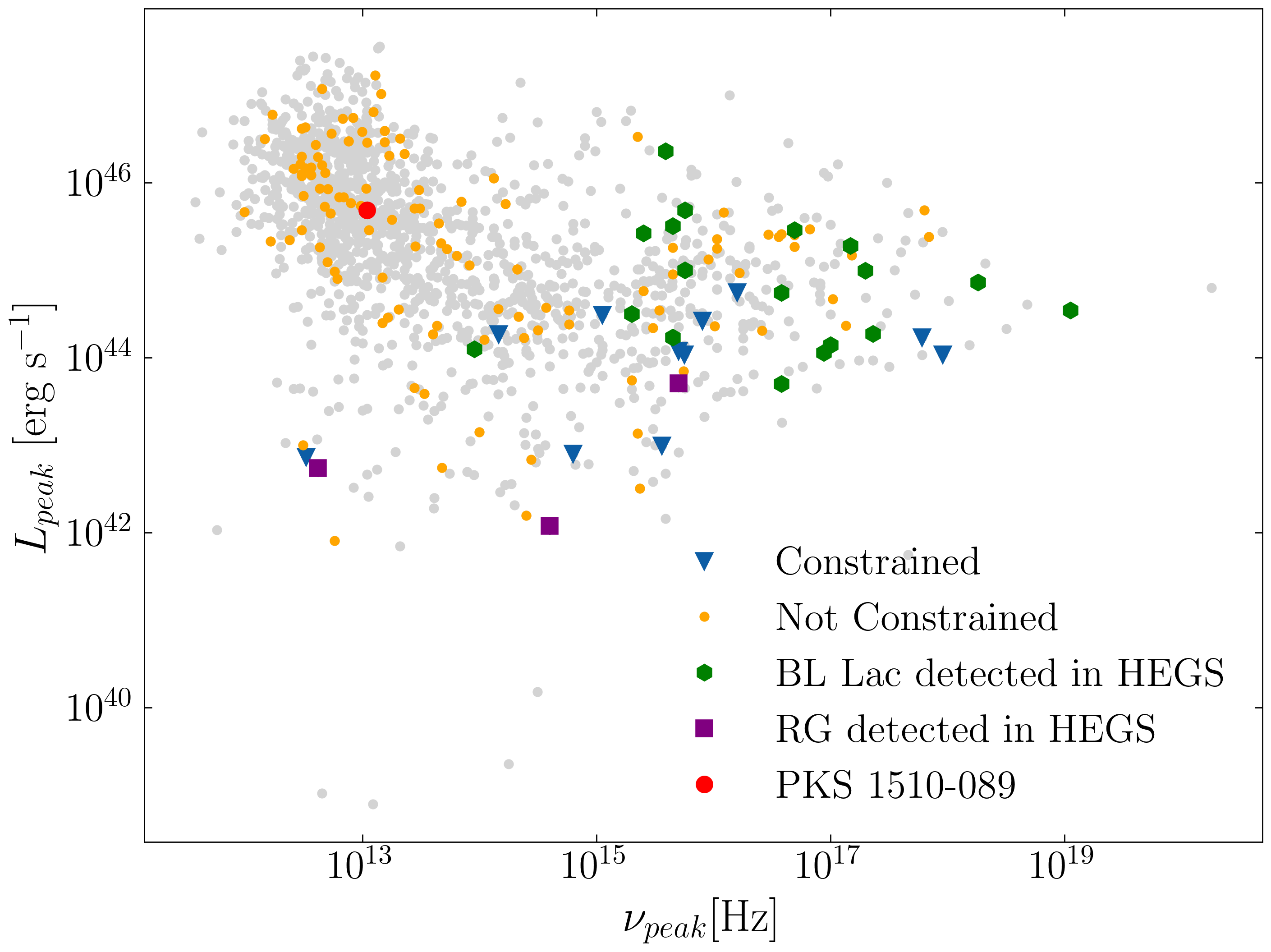}
    \caption{Luminosity at the synchrotron peak, $L_{\rm peak}$, as a function of the synchrotron peak frequency value, $\nu_{\rm peak}$. The data (grey points)  
    were extracted from the BZ Catalogue \citep{2015Ap&SS.357...75M}. Detected sources (BL Lac, RG and FSRQ) in the HEGS FoV are given by green hexagons, purple squares, and a red dot, constrained sources (see Sect.~\ref{sec:Fermicat}) by blue triangles, and not constrained source by orange dots. For the constrained sources, 1 object do not have an entry in the BZ Catalogue: PMN J0039-2220.}
    \label{fig:BS}
\end{figure}

Fig.~\ref{fig:BS} shows the synchrotron peak luminosity as a function of the peak frequency. As is seen in the figure, most of the detected HEGS sources are found to lie in the weak jet (low luminosity)/high frequency synchrotron peak region \citep[see][]{2021MNRAS.505.4726K}, corresponding to the HBL subclass of BL~Lac objects. Likewise, most of the sources for which the VHE data constrain the lower energy extrapolated flux (see Sect.~\ref{sec:Fermicat}) are also in this region of the figure. Although likely, in part, due to a preference in the H.E.S.S. scheduling programme to look at HBL objects, such an explanation cannot entirely explain the preference in the detection of this subclass. This can be appreciated from the fact that more than half of the constrained sources (see Sect.~\ref{sec:Fermicat}), in the HEGS FoV, were not the primary focus of the VHE observations (i.e. were not proposed targets).

\subsection{The extragalactic gamma-ray background}

\begin{figure}
\centering
        \includegraphics[width=0.45\textwidth]{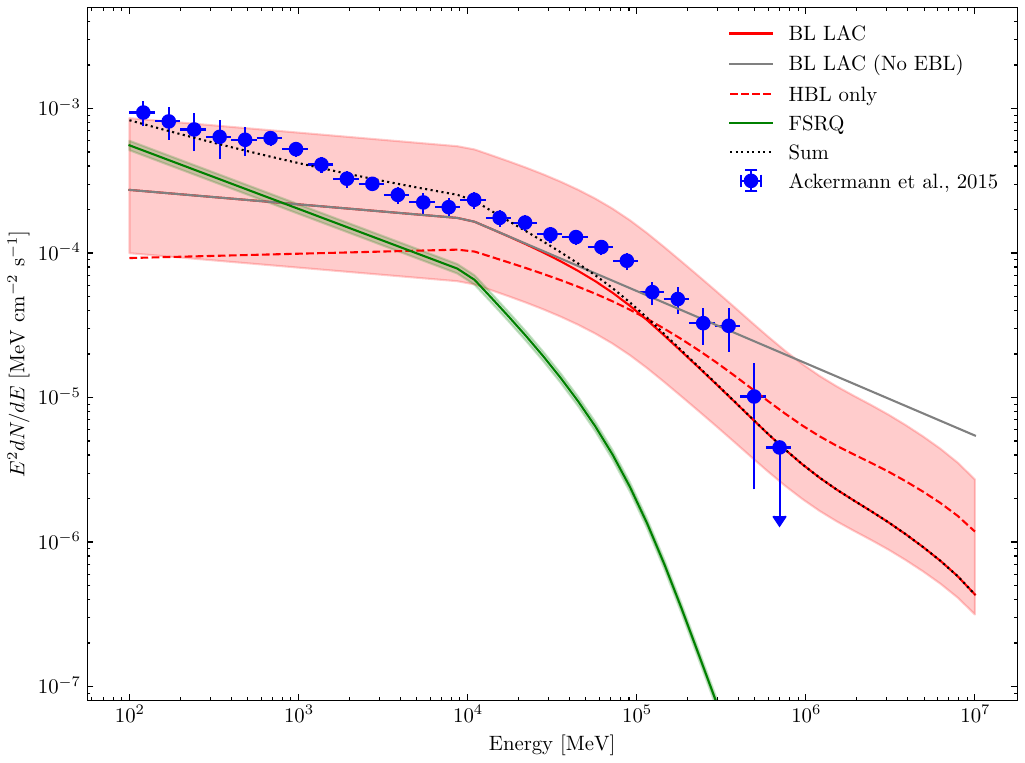}
        
    \caption{EGB measured by \textit{Fermi}-LAT  (blue points). The grey line is the contribution of the BL~Lac objects and red line is the same but with a spectral break $\Delta \Gamma = 0.4$.  The green line is the contribution of FSRQ to the EGB. The red and green areas are the uncertainties for BL~Lac objects and FSRQ, respectively, obtained by only taking the normalisation of the luminosity function into account. The dashed red line was computed using the luminosity function of HBL only derived by \citet{2014ApJ...780...73A}.}
    \label{fig:EGB}
\end{figure}

With the blazar luminosity functions, $\Phi(L_{\gamma},z,\Gamma)$, for both BL~Lac s and FSRQs in hand, an estimation of the total gamma-ray emission from blazars can be obtained. The extragalactic gamma-ray background (EGB) is comprised of the ensemble flux from all VHE-emitting sources. A model estimate of the flux (in ${\rm ph\ cm}^{-2}{\rm s}^{-1}{\rm MeV}^{-1}$) utilising the \textit{Fermi} blazar luminosity function can be obtained by integrating over the blazar luminosity, redshift, and photon index distributions,
\begin{eqnarray}
F_{\gamma}=\int_{L_{\gamma}^{\rm min}}^{L_{\gamma}^{\rm max}}\int_{z_{\rm min}}^{z_{\rm max}}\int_{\Gamma_{\rm min}}^{\Gamma_{\rm max}}\Phi(L_{\gamma},z,\Gamma)e^{-\tau_{\rm EBL}}\frac{dN_\gamma}{dE}\frac{dV}{dz}dL_{\gamma}dz d\Gamma.\ 
\end{eqnarray}
To evaluate $F_{\gamma}$, the same broken power-law as in Sect.~\ref{simu} was used. A break at 10 GeV of $\Delta \Gamma=0.4$ for BL~Lac\ objects and $\Delta\Gamma=0.8$ for FSRQs has then been introduced (see Sect.~\ref{simu}). The outcome from this evaluation of the total flux is shown as the shaded region of Fig.~\ref{fig:EGB}.

As is demonstrated in Fig.~\ref{fig:EGB}, the contribution from
BL~Lac\ blazars dominates the high-energy end of the \textit{Fermi}-LAT  band of the EGB. In contrast to this, the contribution from the FSRQ population dominates at the low-energy end of this band. The dominance of BL~Lac\ objects in the high energy end of the band, can be further broken down into subsidiary components. \citet{2014ApJ...780...73A} derived estimates of the luminosity function of the different sub-classes of BL~Lac objects. While based on lower statistics, such an estimate may offer an improved description the population of HEGS sources. The results of such a breakdown is shown in  Fig.~\ref{fig:EGB} (dashed red line), where it is shown that BL~Lac\ with hard photon indices (aka HBLs) dominate the highest energy end of the EGB, at energies above 100~GeV.

The hard photon index and dominance of the HBL sub-population's contribution to the EGB are of particular relevance and interest for future unbiased blazar
observation programmes such as the planned extragalactic survey by CTAO. 
This survey plans to probe the AGN population in
the extragalactic sky down to 6~mCrab above 125~GeV with 1000 hours of observations.
Such a survey offers the potential to probe both the spectra of FSRQ and BL~Lac objects up
to higher energies than has been previously probed by blazar surveys, providing the 
possibility to further test and improve the spectral description of the luminosity functions currently employed to 
characterise these two blazar populations.

One particular aspect that future observations will help to answer is an improved understanding of the HBL spectral model. The broken power-law description adopted here, consistent with both the \textit{Fermi}-LAT  and H.E.S.S. observations, potentially leads to an over-contribution of HBLs to the EGB, in the VHE band. This potential conflict can be observed in Fig.~\ref{fig:EGB}, in which the HBL model line is above the last upper limit.

Given that the EGB also has contributions from ultra-high-energy cosmic ray (UHECR) cascade \citep{Taylor:2015rla} and neutrino sources \citep{Fang:2022trf}, and with their collective contribution 
to this background increasing with increasing gamma-ray energy, an improved understanding 
of the AGN contribution to the extragalactic gamma-ray background is of great importance. Focusing on the EGB in the energy range $>100$~GeV, of particular importance to this background is the contribution from HBL objects, which provide the dominant contribution to the EGB within this energy range. An improved understanding of the spectral emission from HBL objects, therefore, can potentially provide a useful input in constraining the level of contribution to the EGB from UHECR and neutrino sources.

\section{Summary}
\label{summary}

The H.E.S.S. collaboration has conducted observations of the extragalactic sky over its first decade of operations, during H.E.S.S.-I, with a homogeneous array of four 12-meter Cherenkov telescopes. This large dataset has been partially published over time with different analysis pipelines and instrument calibrations. We have, in this work, re-analysed all the observations made during H.E.S.S.~phase~I in a uniform manner with advanced analysis algorithms and an improved instrument calibration. Using a single common analysis pipeline, a set of extragalactic maps has then been produced, similar to those produced for the Galactic plane in \citet{HGPS_paper}: significance maps, flux and uncertainty of the flux above the analysis energy threshold, upper limit maps and instrument sensitivity maps.

In the analysis, only data for objects considered as extragalactic were re-analysed (removing data with $|b| < 10^{\circ}$ in Galactic coordinates) for a total of  6500\ observation runs for 2720\ hours of data, covering 5.7 \% of the sky. The observation fields, used without information about the target sources, were grouped into clusters. The analysis of these clusters led to the detection of 23 sources, all of which were already known tera-electronvolt emitters. Most of these HEGS objects are AGN, with the noticeable exception of the starburst galaxy NGC~253. Furthermore, within the AGN objects detected, BL~Lac\ objects were the dominant source class. Additionally, two of the AGN detected are RGs, and one is of uncertain sources class (PKS~0625-354). A comparison of the analysis results with the previous H.E.S.S. publication results on these sources was performed. The results obtained here are compatible with the previously published results.

Studies of the flux evolution with time for these sources were carried out. Within the HEGS dataset, seven sources were characterised as variable. Most of these sources (except 1ES~0347-121) were already known to be variable objects at tera-electronvolt energies. All the HEGS sources were found to have a \textit{Fermi}-LAT  counterpart. For the non-variable HEGS objects, we fitted the combined (time-averaged) \textit{Fermi}-LAT  and HEGS SED, within the energy range from 100~MeV up to a few tera-electronvolts. The results from this analysis allowed constraints, such as the high-energy SED peak position, to be obtained on the ensemble of these objects. 

For the majority of the HEGS objects, even after correction for EBL absorption, a different power-law slope was found in the HE and VHE energy ranges, suggesting the presence of a general spectral softening in the spectra of this source population. Evidence for such a spectral break feature was further supported by simulations of the entire HEGS observations, using the \textit{Fermi}-LAT  derived luminosity function. From these simulations, a spectral break was found to be required in order to achieve consistency with the HEGS observational results. A coherent picture for the multi-wavelength spectra of blazars is therefore found, indicating how the ensemble gamma-ray spectrum from the blazar population, derived from catalogues in the HE range, extends up to the VHE range.

The presence of the spectral break in the blazar spectrum, derived from both the \textit{Fermi}-LAT  and HEGS catalogues, was incorporated into the luminosity function for describing the HE/VHE spectral distribution of these objects with luminosity. The expected detection ratio of these two source classes was subsequently estimated using Monte Carlo sampling. The observational bias giving rise to the preference for the detection of BL~Lac\ over FSRQ blazars, was evaluated. This estimated observational bias was found to be consistent with the serendipitous detection rate results of these two source classes. Alongside this evaluation, the expected distribution of spectral index and redshift of BL~lac objects was computed.  Good agreement is found between the actual mean index of the BL Lac objects and the spectral index distribution of the detectable simulated HEGS objects (see comparison of vertical line and blue histogram in the left-panel of Fig.~\ref{fig:simIndexRedshift}).

Furthermore, using the derived broken power-law spectral model to describe the multi-wavelength spectrum of both BL~Lac and FSRQ type source classes, the contribution to the EGB from these objects up to the VHE range was evaluated. A significant contribution to the EGB was found to be expected, with the dominant contribution coming from the HBL sub-class. Furthermore, the possibility that the theoretically estimated EGB from the HBL component may overshoot the model points at even higher energies, close to 1~TeV, was found. Further insights on this tension will be provided by future observations with H.E.S.S. phase II and the forthcoming instrument CTAO.

\begin{acknowledgements}
The support of the Namibian authorities and of the University of Namibia in facilitating the construction and operation of H.E.S.S. is gratefully acknowledged, as is the support by the German Ministry for Education and Research (BMBF), the Max Planck Society, the German Research Foundation (DFG), the Alexander von Humboldt Foundation, the Deutsche Forschungsgemeinschaft, the French Ministry for Research, the CNRS-IN2P3 and the Astroparticle Interdisciplinary Programme of the CNRS, the U.K. Science and Technology Facilities Council (STFC), the IPNP of the Charles University, the Czech Science Foundation, the Polish National Science Centre, the South African Department of Science and Technology and National Research Foundation, the University of Namibia, the National Commission on Research, Science \& Technology of Namibia (NCRST), the Innsbruck University, the Austrian Science Fund (FWF), and the Austrian Federal Ministry for Science, Research and Economy, the University of Adelaide and the Australian Research Council, the Japan Society for the Promotion of Science and by the University of Amsterdam.
We appreciate the excellent work of the technical support staff in Berlin, Durham, Hamburg, Heidelberg, Palaiseau, Paris, Saclay, and in Namibia in the construction and operation of the equipment. This work benefited from services provided by the H.E.S.S. Virtual Organisation, supported by the national resource providers of the EGI Federation.

This work has been done thanks to the facilities offered by the Université Savoie Mont Blanc MUST computing centre.
\end{acknowledgements}

\bibliography{HEGS}
\bibliographystyle{bibtex/aa}

\begin{appendix}
\section{Additional material}

\label{online}

The data that are released together with the present paper consist of a set of FITS files containing the description of the observations, the catalogue, and the maps. 

All the observations in this dataset have been grouped by clusters of observation, as is described in Sect. \ref{ObsMeth}. Each cluster is an independent sky region. The released sky maps were produced independently for each cluster, and each cluster is named according to the sky coordinates of its centre.

The following files are released\footnote{\url{https://hess.science/pages/publications/auxiliary/2024_HEGS/}} : 
\begin{description}
    \item[{\tt HEGS\_FOVs.fits}]: lists the HEGS clusters. This file contains two entries : 
\begin{itemize}
    \item {\tt HEGS\_FOVS} lists the names, positions and extension of each observed FoV. The definition is given in Table~\ref{table:fitsfovs}.
    \item {\tt HEGS\_FOVS\_NIGHTLY\_LIVETIMES} lists the observed live time per night for each cluster. We note that this is only indicative. Since a FoV can be extended on the sky, some observation time in a night for a given field-of-view doesn't necessarily mean that the whole area of the cluster has been observed on that particular night. The definition is given in Table~\ref{table:fitsfovstimes}.
\end{itemize}

\item[{\tt HEGS\_catalogue.fits}]: The HEGS catalogue, which contains several entries:
\begin{itemize}
    \item {\tt HEGS\_sources}: lists the main catalogue results with one source per row. The definition is given in Table~\ref{table:fitsSource}.
    \item {\tt Association}: Association of HEGS source with one source per row. The definition is given in Table~\ref{table:asso}.
    \item {\tt Constrained sources}: \textit{Fermi}-LAT sources within the HEGS FoV that are constrained by the H.E.S.S. observations (see Sect.~\ref{sec:Fermicat}). One source per row. The definition is given in Table~\ref{table:cons}.
\end{itemize}

\item[{\tt Maps}]: For each cluster, a set of maps is released in one FITS file, named "HEGS\_JHHmm±DDd\_Maps\_Loose\_Index3.fits.gz" from the cluster name as listed in the {\tt HEGS\_FOVs.fits} file. The definition is given in Table~\ref{table:fitsMaps}.

\item[{\tt HEGS\_NotConstrainedSources.fits}]: \textit{Fermi}-LAT sources in the FoV of HEGS that are not constrained by the H.E.S.S. observations (see Sect.~\ref{sec:Fermicat}). The definition is the same as the {\tt Constrained sources} one described in Table~\ref{table:cons}.
\end{description}

\begin{table*}
\small{
\caption{Description of the FITS table containing information on the observed FoVs.}
\label{table:fitsfovs}

\begin{center}
\begin{tabular}{lll }
\hline\hline 
Column & Units & description\\\hline

FOV\_Name &  & Name\\
Center\_RA & deg & Right Ascension (J2000) of the centre of the cluster\\
Center\_DEC & deg & Declination (J2000) of the centre of the cluster\\
ExtX & deg & Half-extension of the cluster map along the Right Ascension axis\\
ExtY & deg & Half-extension of the cluster map along the Declination axis\\
\hline 

\end{tabular}
\end{center}
}
\end{table*}

\begin{table*}
\small{
\caption{Description of the FITS table containing information on the observation time per FoV.}
\label{table:fitsfovstimes}

\begin{center}
\begin{tabular}{lll }
\hline\hline 
Column & Units & description\\\hline

FOV\_Name &  & Name\\
Time\_MJD & d & List of start dates (at noon before the night)\\
LiveTime & s & List of observation live time (in seconds) for each night containing observations\\
\hline 

\end{tabular}
\end{center}
}
\end{table*}

\begin{table*}
\small{
\caption{Description of the catalogue FITS table. PL stands for Power Law and LP for LogParabola.}
\label{table:fitsSource}

\begin{center}
\begin{tabular}{lll }
\hline\hline 
Column & Units & description\\\hline

Source\_Name &  & Source name (HESS JHHmm$\pm$DDd identifier)\\
RAJ2000 & deg & Right Ascension (J2000)\\
DECJ2000 & deg & Declination (J2000)\\
nON &  & Number of events in the ON region\\
nOFF &  &Number of events in the OFF regions \\
Alpha &  & Normalisation of the OFF regions\\
LiMa\_Significance &  & Detection Significance\\
LiveTime & h  & Live time in hours \\
Flux\_Spec\_PL\_Diff\_Pivot & cm$^{-2}$ s$^{-1}$ TeV$^{-1}$  & Differential flux at pivot energy\\
Flux\_Spec\_PL\_Diff\_Pivot\_Err & cm$^{-2}$ s$^{-1}$ TeV$^{-1}$   & Statistical error (1 sigma) on Flux\_Spec\_PL\_Diff\_Pivot\\
Index\_Spec\_PL &  & Spectral index \\
Index\_Spec\_PL\_Err &  & Statistical error (1 sigma) on Index\_Spec\_PL \\
Pivot\_Energy & TeV & Pivot energy\\
Spectral\_Model &  & Preferred spectral model\\
Emin\_Spectrum &   TeV& Minimal energy of the spectrum\\
Emax\_Spectrum &  TeV & Maximal energy of the spectrum\\
Flux\_Spec\_LP\_Diff\_Pivot &  cm$^{-2}$ s$^{-1}$ TeV$^{-1}$  & Differential flux at pivot energy\\
Flux\_Spec\_LP\_Diff\_Pivot\_Err &  cm$^{-2}$ s$^{-1}$ TeV$^{-1}$  & Statistical error (1 sigma) on Flux\_Spec\_LP\_Diff\_Pivot\\
Alpha\_Spec\_LP &  & Spectral index \\
Alpha\_Spec\_LP\_Err &  &Statistical error (1 sigma) on Alpha\_Spec\_LP  \\
Beta\_Spec\_LP &  & Curvature value \\
Beta\_Spec\_LP\_Err &  & Statistical error (1 sigma) on Beta\_Spec\_LP\_Err \\
Flux\_Spec\_PL\_Diff\_Pivot\_EBL & cm$^{-2}$ s$^{-1}$ TeV$^{-1}$  & Differential flux at pivot energy\\
Flux\_Spec\_PL\_Diff\_Pivot\_Err\_EBL & cm$^{-2}$ s$^{-1}$ TeV$^{-1}$  & Statistical error (1 sigma) on Flux\_Spec\_LP\_Diff\_Pivot\_EBL\\
Index\_Spec\_PL\_EBL &  & Spectral index, EBL corrected\\
Index\_Spec\_PL\_Err\_EBL &  &Statistical error (1 sigma) on Index\_Spec\_PL\_EBL\\
Pivot\_Energy\_EBL & TeV & Pivot energy \\
Redshift &  & Redshift of the source\\
Energy & TeV  & centre of the energy bin\\
Delta\_Energy\_Up &  TeV  & high energy bound of the energy bin \\
Delta\_Energy\_Low & TeV  & low energy bound of the energy bin \\
Differential\_Flux & cm$^{-2}$ s$^{-1}$ TeV$^{-1}$  & differential flux at Energy \\
Differential\_Flux\_Err & cm$^{-2}$ s$^{-1}$ TeV$^{-1}$ & Statistical error (1 sigma) on Differential\_Flux, 0 if upper limit \\
Time\_MJD & d & Time in days of the Light curve point (only for sources identified as variable) \\
Flux\_above\_300GeV& cm$^{-2}$ s$^{-1}$ & Integral flux above 300 GeV\\
Flux\_above\_300GeV\_Err& cm$^{-2}$ s$^{-1}$ &Statistical error (1 sigma) on Integral flux above 300 GeV\\
\hline 

\end{tabular}
\end{center}
}
\end{table*}

\begin{table*}
\small{
\caption{Description of the association FITS table.}
\label{table:asso}

\begin{center}
\begin{tabular}{ll }
\hline\hline 
Column  & description\\\hline

Source\_Name   & Source name (HESS JHHmm±DDd identifier)\\
Association  & Name of the associated source\\
HEGS FoV  & Name of the HEGS FoV\\
Redshift & value of the redshift\\
Type & Classification of the object\\

\hline 

\end{tabular}
\end{center}
}
\end{table*}

\begin{table*}
\small{
\caption{Description of the FITS table containing information on the constrained sources.}
\label{table:cons}

\begin{center}
\begin{tabular}{lll }
\hline\hline 
Column & Units & description\\\hline

4FGLName &  & Source name (4FGL JHHmm$\pm$DDd identifier)\\
RAJ2000 & deg & Right Ascension (J2000)\\
DECJ2000 & deg & Declination (J2000)\\
Redshift&  & value of the redshift\\
LiMa\_Significance &  & Detection Significance\\
LiveTime & h  & Live time in hours\\
UpperLimit\_1TeV & cm$^{-2}$ s$^{-1}$ TeV$^{-1}$  & Upper limit on the differential flux at 1 TeV (95\% CL)\\
Energy\_threshold & TeV & Average energy threshold\\
\hline 

\end{tabular}
\end{center}
}
\end{table*}

\begin{table*}
\small{
\caption{Description the FITS files containing the maps.}

\label{table:fitsMaps}
\begin{center}
\begin{tabular}{lll }
\hline\hline 
Maps & Units & description\\\hline
LiveTime & h & Acceptance corrected live time in hours \\
SignificanceMap & - & Significance of the signal \\
Average\_EThreshold & TeV & Average energy threshold\\
FluxMap & cm$^{-2}$ s$^{-1}$ TeV$^{-1}$ & Differential flux at 1 TeV \\
FluxErrorMap & cm$^{-2}$ s$^{-1}$ TeV$^{-1}$ & Error on the differential flux at 1 TeV \\
FluxULMap & cm$^{-2}$ s$^{-1}$ TeV$^{-1}$ & Upper limit on the differential flux at 1 TeV (95\% CL) \\
FluxSensitivityMap\_Sig5p0 & cm$^{-2}$ s$^{-1}$ TeV$^{-1}$ & Flux sensitivity at $5.0$ sigma \\
FluxSensitivityMap\_Sig5p7 & cm$^{-2}$ s$^{-1}$ TeV$^{-1}$ & Flux sensitivity at $5.7$ sigma \\
\hline 

\end{tabular}
\end{center}
}
\end{table*}

\end{appendix}

\end{document}